\documentclass[aip,jap,reprint]{revtex4-1}

\usepackage{amsmath}
\usepackage[utf8]{inputenc}
\usepackage{academicons}
\usepackage{xcolor}
\DeclareUnicodeCharacter{0087}{\color{red}{BAD!!}}
\usepackage{graphics}
\usepackage{graphicx}
\usepackage[T1]{fontenc} 

\usepackage[breaklinks=true]{hyperref}
\usepackage{tikz}
\usepackage[all]{hypcap}
\usepackage{orcidlink}
\definecolor{JAPblue}{rgb}{0, 0.682, 0.937}
\hypersetup{
	colorlinks = true,
	linkcolor = {JAPblue},
	citecolor = {JAPblue},
	urlcolor = {JAPblue},
}

\usepackage{upgreek}
\usepackage{rotating}

\begin{document}

\title{Femtosecond temperature measurements of laser-shocked copper deduced from the intensity of the x-ray thermal diffuse scattering}

\author{J.S. Wark~\orcidlink{0000-0003-3055-3223}}\email{justin.wark@physics.ox.ac.uk}
\affiliation{Department of Physics, Clarendon Laboratory, University of Oxford, Parks Road, Oxford OX1 3PU, UK\looseness=-1}

\author{D.J. Peake\orcidlink{0000-0002-5992-6954}}
\affiliation{Department of Physics, Clarendon Laboratory, University of Oxford, Parks Road, Oxford OX1 3PU, UK\looseness=-1}

\author{T. Stevens~\orcidlink{0009-0006-8355-3509}}
\affiliation{Department of Physics, Clarendon Laboratory, University of Oxford, Parks Road, Oxford OX1 3PU, UK\looseness=-1}

\author{P.G. Heighway~\orcidlink{0000-0001-6221-0650}}\email{patrick.heighway@physics.ox.ac.uk}
\affiliation{Department of Physics, Clarendon Laboratory, University of Oxford, Parks Road, Oxford OX1 3PU, UK\looseness=-1}

\author{Y. Ping~\orcidlink{0000-0002-4879-9072}}
\affiliation{Lawrence Livermore National Laboratory, Livermore, CA 94550, USA\looseness=-1}

\author{P. Sterne~\orcidlink{0000-0002-6398-3185}}
\affiliation{Lawrence Livermore National Laboratory, Livermore, CA 94550, USA\looseness=-1}


\author{B. Albertazzi}\affiliation{Ecole Polytechnique, Palaiseau, Laboratoire pour l'utilisation des lasers intenses (LULI), CNRS UMR 7605 Route de Saclay 91128 PALAISEAU Cedex, France\looseness=-1}

\author{S.J. Ali~\orcidlink{0000-0003-1823-3788}}\affiliation{Lawrence Livermore National Laboratory, Livermore, CA 94550, USA\looseness=-1}

\author{L. Antonelli~\orcidlink{0000-0003-0694-948X}}\affiliation{University of York, School of Physics, Engineering and Technology, Heslington York YO10 5DD, UK\looseness=-1}

\author{M.R. Armstrong~\orcidlink{0000-0003-2375-1491}}\affiliation{Lawrence Livermore National Laboratory, Livermore, CA 94550, USA\looseness=-1}

% ...

\date{\today}

\begin{abstract}

We present 50-fs, single-shot measurements of the x-ray thermal diffuse scattering (TDS) from copper foils that have been shocked via nanosecond laser-ablation up to pressures above $\sim$135~GPa. We hence deduce the x-ray Debye-Waller (DW) factor, providing a temperature measurement. The targets were laser-shocked with the DiPOLE 100-X laser at the High Energy Density (HED) endstation of the European X-ray Free-Electron Laser (EuXFEL). Single x-ray pulses, with a photon energy of 18~keV, were scattered from the samples and recorded on Varex detectors. Despite the targets being highly textured (as evinced by large variations in the elastic scattering), and with such texture changing upon compression, the absolute intensity of the azimuthally averaged inelastic TDS between the Bragg peaks is largely insensitive to these changes, and, allowing for both Compton scattering and the low-level scattering from a sacrificial ablator layer, provides a reliable measurement of $T/ \Theta_D^2$, where $\Theta_D$ is the Debye temperature. We compare our results with the predictions of the SESAME 3336 and LEOS 290 equations of state for copper, and find good agreement within experimental errors.  We thus demonstrate that single-shot temperature measurements of dynamically compressed materials can be made via thermal diffuse scattering of XFEL radiation.

\end{abstract}

\maketitle

\section{Introduction}

The dynamical compression of condensed matter on nanosecond timescales, with the pressure applied via laser ablation of a surface layer, provides a means to reach pressures far greater than those that can be applied statically in diamond anvil cells (DACs)~\cite{Coppari2013,Smith2014,Lazicki2021}.  Tailoring of the temporal profile of the applied optical laser radiation can lead to samples either being shock compressed to states along the Hugoniot, or ramped more slowly to the high-pressure state, keeping the sample cooler, and closer to the isentrope.  Subsequent pulsed x-ray diffraction provides structural information on a timescale short compared with that of the nanosecond compression~\cite{Wark2022}. This combination of laser-compression and pulsed diffraction has been applied at a number of high-power--laser, synchrotron, and XFEL facilities over the past few decades, providing a wealth of information on high strain-rate deformation physics and phase transitions at pressures from a few GPa to well into the multi-TPa regime~\cite{Wark1987,Wark1989,Loveridge-Smith2001,Turneaure2007,Turneaure2007b,Rygg2012,Suggit2012,Milathianaki2013,Lazicki2015,Gleason2015,Denoeud2016,Wang2016,Turneaure2016,Wehrenberg2017,Polsin2018,Wicks2018,Sharma2020}, which comprises a region of parameter space of relevance to the physics of planets both within our own solar system and beyond~\cite{Seager2007,Jeanloz2007,Valencia2009,Pickard2010,Coppari2013,Smith2014,Helled2017,Duffy2019}. 

Whilst density information can be provided by diffraction, pressure can be deduced via interferometric measurements of the velocity of an interface within, or the free surface of, the target via the VISAR technique (Velocity Interferometer System for Any Reflector)~\cite{Barker1972,Celliers2004}.  Temperature, however, has proven to be a more difficult parameter to measure within such dynamical compression experiments, particularly when the temperatures are too low to be extracted from pyrometric techniques on such short timescales~\cite{Hereil2000,Brantley2021}.  One method that has proven successful at the Omega laser and at the National Ignition Facility (NIF) is  EXAFS (Extended X-ray Absorption Fine Structure), where the depth of modulations in the x-ray absorption coefficient above a K- or L-edge is sensitive to the Debye-Waller (DW) factor (effectively proportional to $T/ \Theta_D^2$, where $\Theta_D$ is the Debye temperature)~\cite{Ping2013,Ping2016,Sio2023}.  If we assume that $\Theta_D$ as a function of compression can be calculated reliably, 
or inferred by other local sound-velocity measurements, then this method provides a means to extract temperatures. Alternatively, as has been done recently, the EXAFS signal can be compared directly with spectra produced from molecular dynamics simulations, bypassing the need to quote a specific value for the Debye temperature~\cite{Sio2023}. 

Although EXAFS measurements have been proven to provide temperature information on the facilities cited above, they are only made possible by the fact that the very large optical laser energies available at them allow for the creation of a separate, bright, short (subnanosecond), broad-band and spectrally structureless laser-plasma-based diverging source of x-rays, which are subsequently absorbed by the dynamically compressed target, and the resultant absorption spectrum recorded.  Such a source is not readily available at x-ray FEL facilities, where a significant proportion of such compression studies are now performed.  FEL facilities have the advantages provided by the highly monochromatic nature of the x-ray beam and its ultrashort pulse-length (which results in the x-ray source being considerably brighter than those provided by a laser-produced plasma).  As a result, other methods of using the x-rays to provide a temperature measurement at FELs have been sought.

One obvious candidate for a temperature measurement is the DW effect as applied to the elastic scattering, whereby the ratios of the intensities of the Bragg peaks  are used to deduce $T |{\bf G}|^2/ \Theta_D^2$, where $\bf{G}$ is the reciprocal lattice vector associated with the Bragg reflection of interest. Whilst in EXAFS the DW effect reduces the depths of the modulations in the absorption coefficient above an absorption edge as the DW factor increases, in diffraction the intensity of  the higher-order Bragg peaks decreases compared with those of lower order,  and indeed the total elastic scattering decreases with a concomitant increase in the TDS.  In both cases the underlying physics is related to the  thermally induced deviation of atoms from their perfect-lattice positions. For diffraction, this introduces a degree of randomization of the phase of the x-rays  scattered from each atom, whereas in the case of EXAFS the phase of the ejected and re-scattered photo-ejected-electron is influenced (note there are thus slight differences in the two DW factors, as EXAFS is probing short-range order, whilst diffraction probes on longer lengthscales).

However, under the influence of dynamic compression, significant texture changes may take place within the sample, making such DW measurements via elastic diffraction difficult: the relative intensities of the Bragg peaks are heavily influenced by the overall orientation distribution function (ODF) of the grains within the sample, which itself changes owing to plastic flow (an issue that does not affect the EXAFS technique, as the absorption is independent of texture). Although simulations under elastic compression seem to indicate the technique might have some merit~\cite{Murphy2008}, previous experimental attempts to deduce DW factors from the elastic scattering from shocked samples probed with short pulses of x-rays of synchrotron radiation have proven to be unsuccessful~\cite{Sharma2021}, and it has been posited that the copious defects that are produced under shock compression may also influence the reliability of this approach. Indeed, within the measurements we present here, we have found that the Bragg-peak elastic scattering cannot be used to reliably extract DW factors owing to texture, indicating that such an approach might only be feasible in situations where the target is largely free from texturing effects (which may be the case, for example, if its thermodynamic path has taken it through into the melt, with subsequent refreezing).

Given the difficulties associated with measuring the effects of the DW factor on the intensity of the Bragg peaks, it has recently been suggested that temperatures in such experiments at FELs could be obtained via spectrally resolved inelastic x-ray TDS from the phonons within the compressed sample~\cite{McBride2018,Descamps2020a,Wollenweber2021}, probing at momentum transfers between Bragg peaks. If such a method were feasible, it would have the advantage that the temperature measurement would rely solely on the principle of detailed balance (with no knowledge of the Debye temperature required), whereby the temperature is inferred merely from the ratio of the intensities of the Stokes and anti-Stokes peaks. However, given the thermal phonons within the compressed samples have maximum energies of just a few 10's of meV, yet the incident x-rays are of order 10's of keV, such experiments require an extremely high degree of monochromaticity in both the x-ray beam ($\lambda / \Delta \lambda > 10^{6}$) and associated light-collecting spectrometer, which consequently make them extremely photon hungry.  Indeed, the scattering cross sections are such that with current total FEL x-ray energies of order a millijoule per pulse, it is likely that many hundreds, if not thousands, of identical shots would be required to build up sufficient signal to make a temperature measurement, even with narrow-band spectral seeding of the FEL beam, precluding single-shot temperature measurements such as are afforded by the EXAFS technique.  Furthermore, as temperatures start to exceed the Debye temperature, the ratio of the Stokes and anti-Stokes components approaches unity, severely limiting the materials and range of temperatures over which this technique can be usefully employed.

\begin{figure*}
    \includegraphics{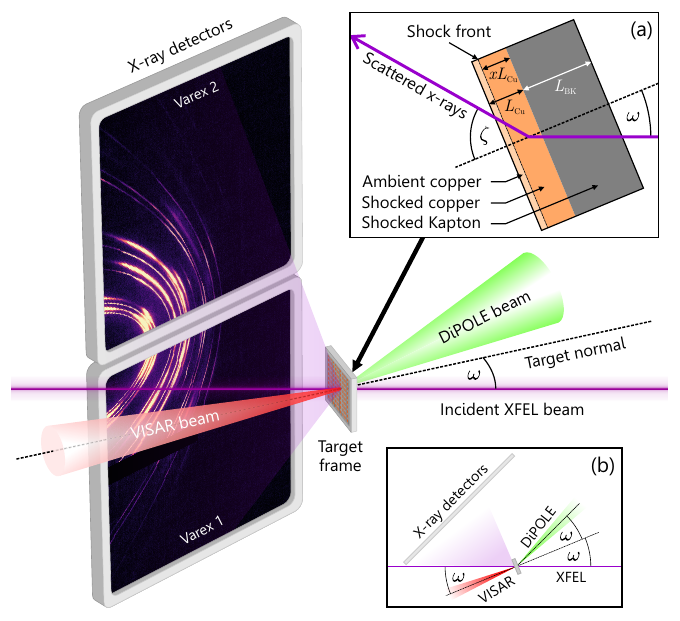}
    \caption{Experimental setup at the High Energy Density (HED) scientific instrument. Ablatively driven shock waves are launched using 10~ns pulses of frequency-doubled radiation from the DiPOLE 100-X laser into targets comprising a Kapton-B ablator of thickness $L_{\mathrm{BK}} = 50$~$\upmu$m glued to a copper foil of thickness $L_{\mathrm{Cu}} = 25$~$\upmu$m. Targets are probed before shock breakout with a beam of 18~keV photons from the x-ray free-electron laser (XFEL) directed at angle $\omega=22.5^\circ$ to the target normal. Resulting diffraction patterns are recorded on a pair of downstream Varex detectors placed symmetrically above and below the beam path. The targets' rear-surface velocity history is measured using a two-leg Velocity Interferometer System for Any Reflector (VISAR). (a) Close-up of x-ray path through partially compressed target. An x-ray incident at angle $\omega$ to the target normal and scattered into angle $\zeta$ traverses a shocked Kapton layer, a shocked Cu layer, and an ambient Cu layer, the latter having thickness $(1-x)L_{\mathrm{Cu}}$, where $x$ is the mass fraction of the Cu traversed by the shock. (b) Simplified top-down view of the experimental setup, illustrating directions of the incident beams and shadowing of scattered x-rays by the target.}
    \label{F:setup}
\end{figure*}

It is in the above context that we demonstrate here that the absolute intensity of the spectrally unresolved (but resolved in scattering angle) TDS between the Bragg peaks can provide a reliable measure of $ T/ \Theta_D^2$.  As no spectral resolution whatsoever is required, such measurements can easily be made on a single-shot basis.  Furthermore, we show  that this inelastic scattering intensity, when averaged over a reasonable range of azimuthal angles, is much more robust against changes in texture than the elastic scattering, and mainly depends on changes in the DW factor.  This TDS signal, at least for the mid-$Z$ target of Cu studied here, also dominates over both Compton scattering from the Cu, and the scattering (elastic and Compton) from the low-$Z$ ablation layer frequently used in such experiments (all of which can, furthermore, be taken into account in the analysis procedure).  

We present results from laser-shocked Cu up to specific volume ratios $V/ V_0$ of 0.7 (where $V_0$ is the specific volume of the ambient material), corresponding to pressures (according to the SESAME 3336 EOS) of order 137~GPa.  The intensity of the TDS is compared with predictions of a simple model, based on the classic work of Warren~\cite{Warren1951,Warren1953}.  When we adapt the Warren model to take texture effects into account, we find negligible differences for the azimuthally averaged TDS between highly textured samples and random powders, demonstrating the applicability of the original simple Warren model to the TDS scattering (the same statement does not apply for the elastic Bragg scattering).  Applying this model to the experimental data, we extract values of $T/ \Theta_D^2$ along the Hugoniot.  We compare our results with the predictions of the LLNL LEOS 290 and SESAME 3336 EOS, both of which provide values for $\Theta_D$ and $T$ along the Hugoniot.  We also make comparison with the results of the historical shock compression experiments by Al'tshuler and co-workers, where temperatures were deduced from a Mie-Gr{\"u}neisen model~\cite{Altshuler1960}.  Within the experimental error of our measurements, we find broad agreement with these models, thus demonstrating the feasibility of using single shot TDS as a temperature measurement for dynamically compressed matter.

The paper is laid out in the following manner.  In section \ref{S:Setup} we outline the experimental set-up. Then, in section \ref{S:Results} we present the experimental results, and show how they compare with simulations, thus allowing an extraction of the DW factor (and hence temperature if we assume a knowledge of $\Theta_D$ under compression).  We compare our results with those predicted by the models referenced above. Finally, in section \ref{S:Discussion} we discuss the results, the potential advantages and  limitations of the technique, and comment upon ways whereby more accurate measurements of the DW factor could be made in the future.

\section{Experimental Set-Up} \label{S:Setup}

The dynamic-compression experiment was performed in Interaction Chamber 2 (IC2) of the High Energy Density (HED) scientific instrument at the European X-ray Free-Electron Laser (EuXFEL). We show the configuration of the target chamber in \hyperref[F:setup]{Fig.~1}.

To shock-compress our targets, we used the recently commissioned \cite{Gorman2024} DiPOLE 100-X laser system \cite{Mason2018}. Targets were irradiated with 10~ns pulses of frequency-doubled (515~nm) light containing up to 40~J of energy, concentrated into a drive spot of either 500~$\upmu$m or 250~$\upmu$m diameter depending on the desired pressure. For the 500-$\upmu$m drive spot -- which allowed access to values of $V/V_0$ of just below 0.75 -- we used a flat-top (constant intensity) laser pulse; for the very highest-pressure shots ($V/V_0=0.7$), driven using a smaller 250-$\upmu$m drive spot, the laser intensity was linearly ramped by 10\% over the course of the 10~ns pulse duration to prevent the decay of the ablation pressure. The laser energy was monitored by siphoning off a small portion of the main beam into a calorimeter situated outside the interaction chamber.

The targets comprised a 50-$\upmu$m-thick polyimide (Kapton~B, DuPont) ablator layer glued to a 25-$\upmu$m-thick rolled Cu foil (Goodfellow). Targets were diced into $5\times5$~mm\textsuperscript{2} tokens and mounted in the recesses of a ladder-type frame, which were separated from one another by at least 10~mm. To ensure repeatability, all targets were mounted with a consistent orientation, such that their rolling direction (RD) was vertical to within a few degrees. We confirmed the consistency of the target orientations by comparing their `pre-shots' (diffraction patterns obtained on the ambient target prior to shock-compression), and verifying that the azimuthal structure in their Debye-Scherrer rings was compatible with a single underlying crystallographic texture.  For the rolled foils used in this experiment, the dominant component of orientation distribution function (ODF) was largely consistent with a $\beta$-fiber texture, which is often seen in such copper samples.

Our primary diagnostic was femtosecond x-ray diffraction. We illuminated the shock-compressed targets with 50~fs bursts of 18~keV x-rays traveling at angle $\omega=22.5^\circ$ to the target normal and coincident with the center of the optical drive spot. The x-ray spot size was set to 45~$\upmu$m for shots taken with the larger 500~$\upmu$m drive spot, and reduced to 20~$\upmu$m for higher-pressure shots taken using a 250~$\upmu$m spot. We endeavored to time the x-ray pulse relative to the onset of the drive laser so as to probe the targets just before the shock wave reached the rear surface of the Cu layer. We were generally successful in timing our shots such that the fraction of the Cu layer traversed by the shock [referred to as $x$ in \hyperref[F:setup]{Fig.~1(a)}] was at least 60\%. However, the demands of the inelastic scattering measurement are such that for the most accurate measurements we needed to sift our data for shots for which $x\ge0.8$; this will be addressed further in section~\ref{S:Results}. The shot-to-shot XFEL intensity was measured using an x-ray gas monitor (XGM) 108.8 m upstream of the center of the target chamber, with an absolute measurement accuracy of $~\pm10\%$ \cite{Maltezopoulos} (see Supplementary Material).

X-ray diffraction patterns were recorded on a pair of 4343CT Varex flat-panel detectors. The detectors were placed symmetrically above and below the x-ray beam path at a distance of 225~mm from the target and rotated through $45^{\circ}$ about the vertical, thus giving azimuthal angular coverage over the domain $\varphi\in(-80,80)^\circ$ and polar coverage over $2\theta\in(5,65)^\circ$. Diffraction beyond a scattering angle of $65^\circ$ was generally weak due to self-attenuation from the target itself [see \hyperref[F:setup]{Fig.~1(b)}]. We deduced the detector positions precisely by fitting diffraction patterns from standard powderlike CeO\textsubscript{2} calibrants using the \textsc{dioptas} software package \cite{Prescher2015}.

Whilst the HED instrument does house a two-leg line-imaging VISAR instrument, whose beams independently monitor the motion of the copper layer's rear surface, data collected in our experiment (which was a component part the first experiments performed on this facility by the user community) was of insufficient quality to extract rear-surface velocities from fringe shifts, and the  VISAR instrument was thus principally used to measure shock breakout times, as has been reported elsewhere~\cite{Gorman2024}.  As a result, we shall present our results as a function of the specific volume ratio, $V/V_0$, as determined directly from the x-ray diffraction, and the pressures we quote will be those predicted by the SESAME 3336 equation of state for the associated compression.

\begin{figure*}
    \includegraphics[width=16cm]{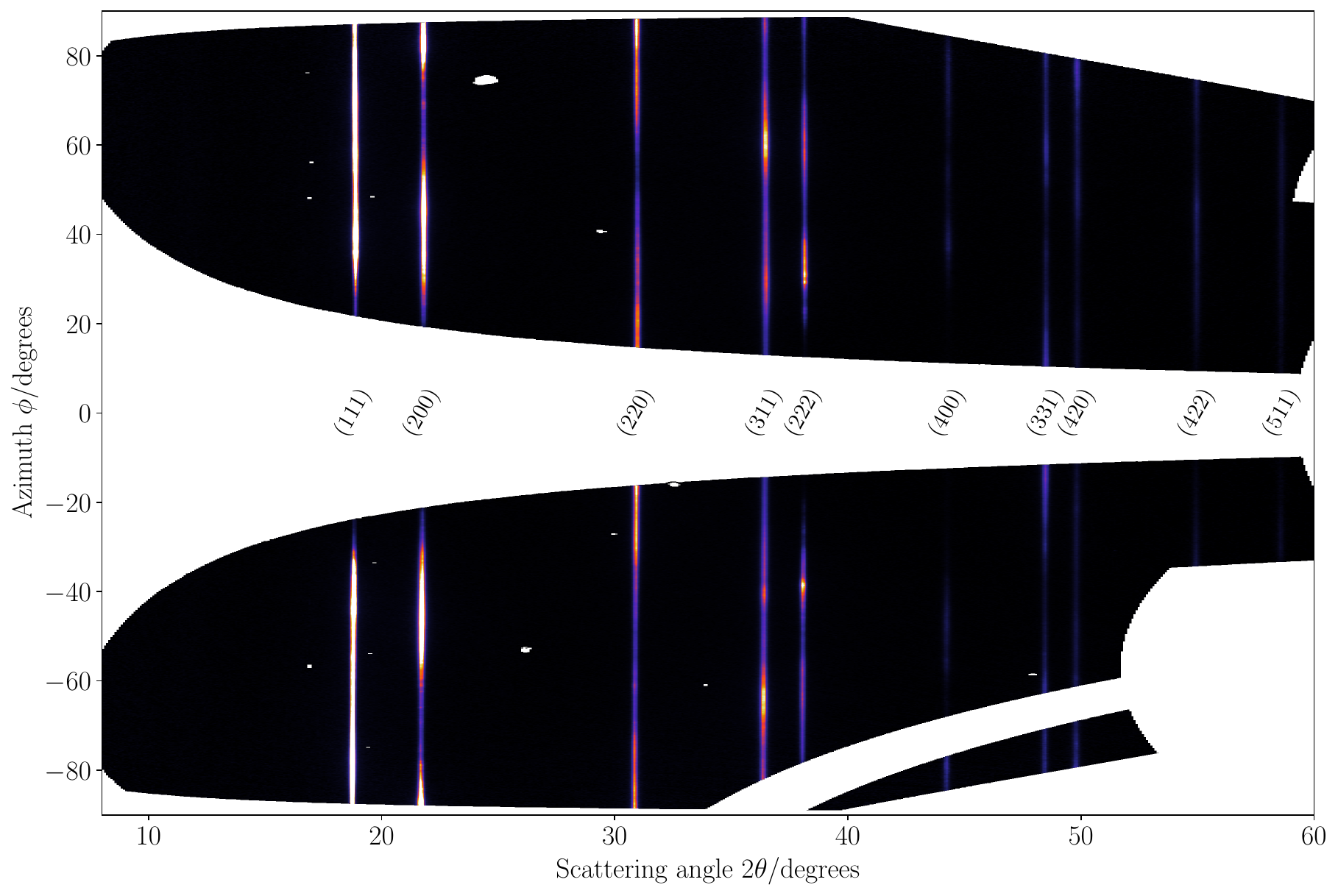}
    \caption{Diffraction data collected on the Varex detectors on an unshocked copper sample. The intensity is corrected for x-ray polarization, pixel solid angle, and the attenuation due to the aluminium filter.}
    \label{F:preshotvarex}
\end{figure*}

\section{Experimental Results}\label{S:Results}

Our aim is to measure the intensity of the angularly resolved  x-ray TDS from the shocked region of the copper sample. As we shall find below, when copper is shocked from ambient conditions to values of $V/V_0$ order 0.7, the strength of the TDS in the regions of interest, between the  Bragg peaks, changes by factors of around two to three, and it is this intensity change that is ultimately a measure of the DW factor, and which we shall also show is insensitive to texture.  A number of effects need to be taken into account in order to achieve this goal with the degree of accuracy which will allow us to infer a meaningful temperature measurement.

Firstly, we need accurate measurements of the incoming x-ray flux on each shot, to which we can normalize the intensity of the diffracted x-rays recorded on the Varex detectors.  Such x-ray flux measurements were made by use of the X-ray Gas Monitor (XGM) discussed in section~\ref{S:Setup}.

Secondly, the largely structureless 50-$\upmu$m thick Kapton ablator layer will scatter over a wide range of angles both due to elastic scattering, and to incoherent Compton scattering, and this combined scattering must be subtracted from the overall experimental signal if only the scattering from the copper is to be considered.  We shall show that owing to the fact that Kapton is of much lower average atomic number, the total scattering from it is weaker than the TDS from the copper. 

Thirdly, at the photon energies used here (18 keV), for Cu the incoherent Compton scattering cross section is non-negligible, and when integrated over all angles has a value approximately 15\% of that of the elastic scattering~\cite{Hubbell1975,Berger1999}.  We will show below that this implies that the Compton scattering from the Cu is still well below the TDS signal, even under ambient conditions, yet is of a level that it should be taken into account in the overall analysis.

Fourthly, as well as x-ray scattering, x-ray absorption is taking place, both while x-rays traverse the target as they propagate along the incident FEL beam direction, and subsequently after they scatter, as they make their way through the target to the detector, as illustrated in \hyperref[F:setup]{Fig.~1(a)}.  This effect can readily be taken into account by using the known absorption coefficient of the target~\cite{Dinnebier2023}.  In addition to this angle-dependent absorption within the target itself, photoelectric absorption in the aluminum filter covering the Varex detectors must also be accounted for, as has been discussed in reports of previous experiments on this facility~\cite{Gorman2024}.

Lastly, we note that not all of the copper target is shocked at the time the diffraction pattern is recorded.  Clearly we would like the vast majority of the target to be in the shocked state, and we need to know what fraction has been shocked (the $x$ in Fig.~\ref{F:setup})  We will show how $x$ can be determined from a measure of the intensity of the diffraction from the thin unshocked layer of Cu at the rear of the target. We will also show that the statistical uncertainty in our measurements decreases markedly for those shots where $x>0.8$.

\begin{figure*}
    \includegraphics[width=16cm]{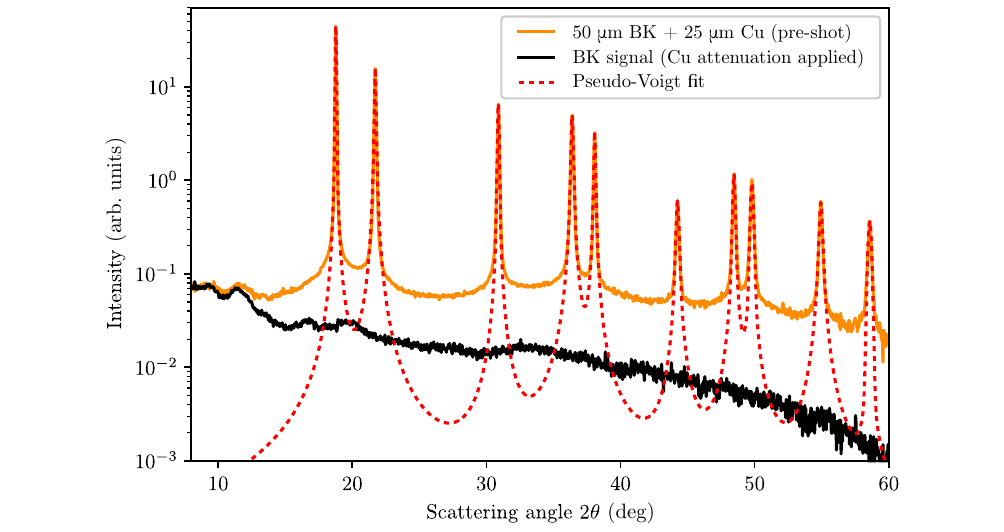}
    \caption{Diffraction signal from an unshocked 25-$\upmu$m thick Cu sample overcoated with 50-$\upmu$m of Kapton, and, on the same scale, the diffraction signal from 50-$\upmu$m Kapton with the x-ray attenuation due to the copper applied. Also shown is the sum of Voigt-profile fits to the Bragg peaks of the Cu sample.}
    \label{F:preshotdata}
\end{figure*}

\begin{figure*}
\includegraphics[width=16cm]{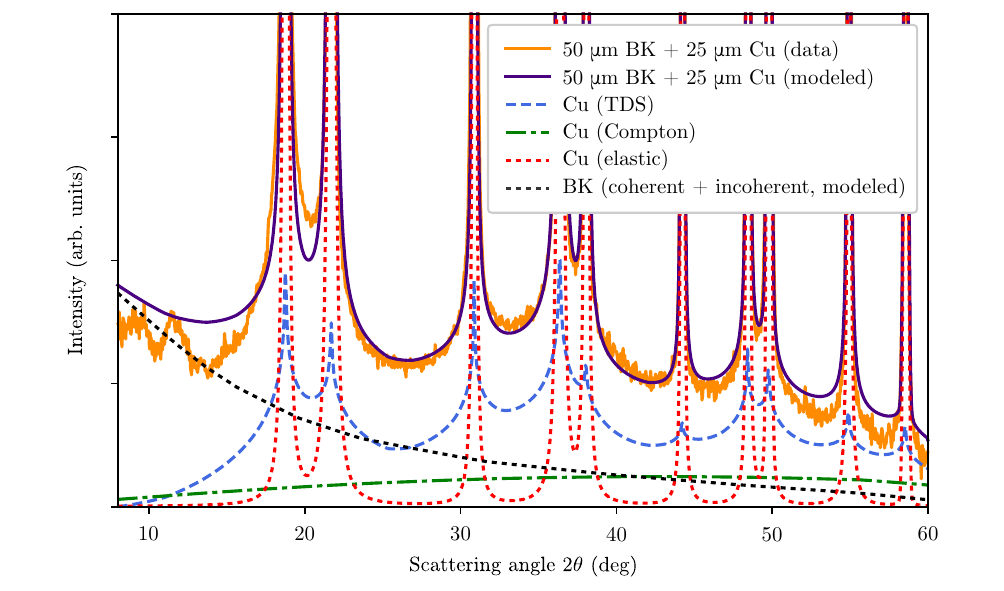
}
    \caption{Total simulated diffraction from an unshocked 25-$\upmu$m thick Cu sample overcoated with 50 $\upmu$m of Kapton, breaking the signal down into elastic, thermal diffuse, and incoherent (Compton) scattering produced by the copper and the total scattering produced by the Kapton.  The experimental data is also shown for comparison.}
    \label{F:theoreticalpreshotdata}
\end{figure*}

The initial points mentioned above can be further elucidated by consideration of data obtained from unshocked targets. In \hyperref[F:preshotvarex]{Fig.~2} we show the Varex images of the diffraction from an unshocked target which comprised a Kapton-coated 25-$\upmu$m thick copper foil (as described in the previous section). Raw data from the Varex detectors has been transformed into $(2\theta, \phi)$ space by use of \textsc{dioptas}~\cite{Prescher2015}, which takes into account the effects of polarization and the solid angle subtended to the target by each pixel.  We have also removed the effects of the angle-dependent absorption due to the filter over the detector (but not the effects of absorption within the target itself). It can be seen that we are recording scattering angles $ 2 \theta$ that range from below $10^{\circ}$ up to around $60^{\circ}$, and for ambient copper we can readily observe all diffraction peaks up to the degenerate (333)/(511) reflections.  The azimuthal coverage in the angle $\phi$ is dependent upon the scattering angle, but as can be seen this coverage is large, and in total can extend up to almost $120^{\circ}$.

The azimuthal average of the data from \hyperref[F:preshotvarex]{Fig.~2} is shown in \hyperref[F:preshotdata]{Fig.~3}.  On the same plot we show the diffraction signal from a target that simply comprises 50~$\upmu$m of Kapton.  This signal is also corrected for polarization and pixel solid-angle effects, and the filter over the detector.  However, to enable us to see the relative contribution that the Kapton makes to our Kapton-coated copper data, on this plot we have reduced the intensity of this signal by an amount corresponding to passing through a 25-$\mu$m thick Cu target at the appropriate scattering angles. It can thus  be seen that the total scattering from the Kapton alone is at least a factor of two weaker than the scattering from the target comprising 25-$\upmu$m Cu coated with Kapton over the whole range of scattering angles, save a region below the Cu (111) peak, where it also starts to exhibit some structure.  

Also shown in \hyperref[F:preshotdata]{Fig.~3} is the sum of pseudo-Voigt profile fits to the elastic Cu Bragg peaks. Note from these fits that in the region between the Bragg peaks, the observed additional scattered intensity is not due to the wings of the Bragg peaks, and thus not due to defect-induced peak broadening, but is primarily caused by thermal diffuse scattering. This can be demonstrated by modeling the diffraction from the Cu according to the classic theory of Warren~\cite{Warren1951,Warren1953}, with higher order scattering calculated using the approximation due to Borie~\cite{Borie1961}. 

In Warren's classic theory of TDS (which for the sake of completeness we summarize in the Supplementary Material), in the limit of temperatures comparable to or greater than the Debye temperature, the TDS for a randomly oriented powder sample of a given crystal type (here face-centered-cubic) as a function of $a \sin \theta / \lambda$, where $a$ is the lattice parameter, and $\lambda$ the wavelength of the x-rays, depends on (twice) the DW factor $2M$, which for an element of mass $m$ is given by
\begin{equation}
    2M = \frac{12h^2}{mk_B}\frac{T}{\Theta_D^2}\left(\frac{\sin\theta}{\lambda}\right)^2 \quad .
\end{equation}
Warren also details how to calculate the elastic scattering for individual Bragg peaks~\cite{Warren1953}. Using the simple Warren theory, we show in \hyperref[F:theoreticalpreshotdata]{Fig.~4} the predicted elastic and TDS scattering from a 25-$\upmu$m thick Cu foil including the effects of absorption within the Cu (and assuming, at this stage, random texture -- we will consider the issue of the effects of texture below). Whilst the individual intensities of each of the  experimental Bragg peaks do not quite fit the Warren theory (and this is indeed due to texture), it is clear that there is an excellent fit to the TDS.  We note that whilst TDS has previously been registered at an XFEL, and its increase observed as the sample has been driven into the melt, direct quantitative comparison with the Warren model has not been made~\cite{Hartley2021}.

In \hyperref[F:theoreticalpreshotdata]{Fig.~4} we also show the predicted total scattering from the Kapton sample, where we have assumed that the Kapton is completely structureless (i.e., we simply make the appropriate sum of the squares of the atomic form factors) to calculate the elastic scattering, and we calculate the incoherent (Compton) scattering using data from Hubbell and co-workers~\cite{Hubbell1975}.  Also shown is a calculation of the incoherent (Compton) scattering from Cu (note this starts to fall off slightly at higher scattering angles, due to absorption within the Cu target). \hyperref[F:preshotdata]{Figures~3} and \hyperref[F:theoreticalpreshotdata]{4} together clearly illustrate two important points: we are indeed measuring the inelastic TDS between the Bragg peaks, and that it is several times more intense than both the overall scattering from the Kapton and the incoherent scattering from the Cu (all of which can subsequently be taken into account in our analysis of the TDS).

Before considering data from shocked samples, we now consider briefly the effects of texture. In the Supplementary Material, we describe how we have adapted the classic theory of Warren both for the elastic Bragg scattering and the inelastic TDS to take into account  texture effects.  In the Debye-Scherrer geometry, when viewed in reciprocal space, elastic scattering occurs when the Ewald sphere (of radius the incident k-vector) intersects the Polanyi spheres (the spheres with radii corresponding to the magnitude of the reciprocal lattice vectors of allowed reflections).  In the Warren theory, first-order inelastic TDS can occur at a point on the Ewald sphere by the addition of the wavevector of a phonon to the wavevector of a point on the Polanyi sphere. Warren makes the assumption that the Brillouin zone can be approximated as a sphere, with the radius $q_B$ of this sphere for a face-centered-cubic crystal (and thus the wavevector of the most energetic phonon) given by
\begin{equation}
    q_B = \frac{2\pi}{a}\left(\frac{3}{\pi}\right)^{\frac{1}{3}}\ .
\end{equation}

For a perfectly random powder, the scattering power of a point on a given Polanyi sphere is uniform and proportional to the multiplicity of the reflection, giving rise to uniform Debye-Scherrer elastic scattering rings (as a function of azimuthal angle, and neglecting the Lorentz factor etc.), and results in the standard Warren formula for TDS.  In essence, our texture-dependent modification to the Warren model comprises numerically integrating the contributions to both the elastic scattering and TDS, based on an appropriate weighting of all of the different points on the Polanyi spheres, having calculated those weightings from a given ODF determined by the texture. Importantly, we find that whilst the azimuthally averaged relative intensities of the elastic Bragg peaks are, as expected, significantly modified by texture, this is not the case for the TDS.

\begin{figure}
     \includegraphics[width=0.95\columnwidth]{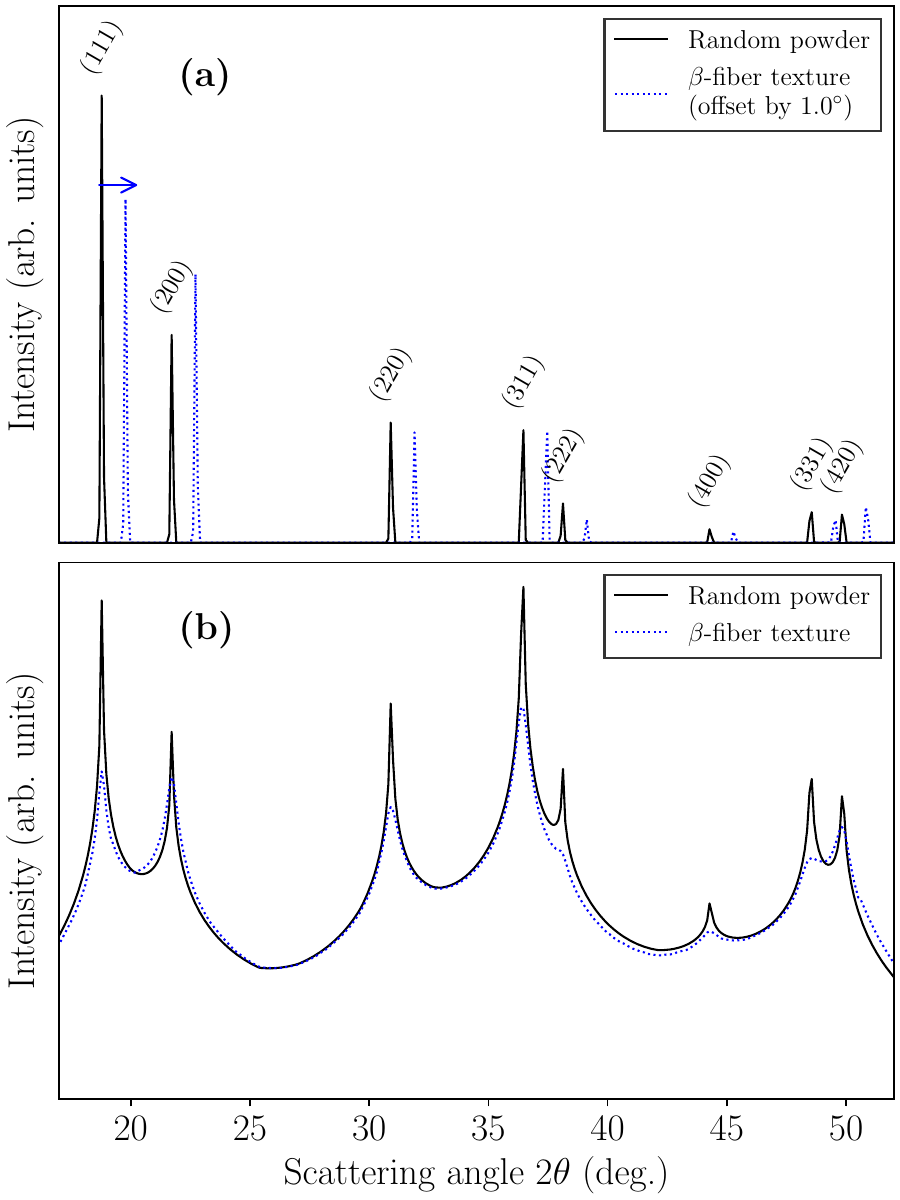}
    \caption{The calculated azimuthally integrated elastic x-ray scattering signal (upper plot) and TDS (lower plot) for an fcc random powder predicted by the analytic solution of Warren \cite{Warren1953} (black). Also shown are results from the present numerical model (blue, dashed) for a $\beta$-fiber-textured polycrystal with surface normal inclined at $22.5^\circ$ to the incident x-rays. The elastic scattering peaks for the fiber-textured case have been offset by 1$^{\circ}$ to enable the intensity differences compared with the powder case to be seen.  Note the insensitivity of the (azimuthally averaged) TDS to the texture.}
    \label{F:warren_powder}
\end{figure}

In order to demonstrate the veracity of the above statement, in \hyperref[F:warren_powder]{Fig.~5} we plot the simulated azimuthally averaged elastic and inelastic TDS scattering from copper, ignoring here the effects of absorption, under ambient conditions for both a perfect powder, and for a $\beta-$fiber sample with a $5^\circ$ spread, where the incident x-ray direction, sample normal, and fiber-axis correspond to those in the experiment, and the azimuthal average has been taken over the same range as that of the experimental data. We choose this texture as the elastic peaks seen in our experimental data are consistent with a large $\beta-$fiber component.  It is clear that the elastic scattering changes considerably due to texture effects -- note in particular the large differences in intensity of the (111) and (200) peaks between the textured and untextured sample -- but, in contrast, the changes to the TDS are negligible, particularly in regions between the Bragg peaks.  Indeed, even if we change the texture (with a simple plasticity model), we find changes in the intensity of the TDS scattering of less than 5\%, a figure which is small compared with the 200\% to 300\% changes in inelastic TDS intensity that we shall find upon shock compression.

The insensitivity of the TDS to texture occurs because for a particular point in reciprocal space, away from the Polanyi spheres, inelastic scattering can occur via the additional wavevector of phonons from all points on the Polanyi sphere that lie within a wavevector of magnitude the Brillouin zone.  As so many points on the Polanyi sphere are thus sampled (albeit with an integral over phonon wavevectors that differs from that of the perfect powder), the effects of nonuniform scattering power on the Polanyi sphere (i.e., texture) is sufficiently smoothed that it is drastically reduced for the TDS.  Indeed, as is well known, inelastic scattering is still observed in this geometry even in the case of a single crystal -- though in that case its distribution throughout reciprocal space would start to exhibit the symmetry of the crystal.  In the case of samples textured to the degree used in this experiment, however, it is clear that the azimuthally averaged TDS differs negligibly from the uniform powder case, allowing us to ignore the effects of texture upon it.  These findings are consistent with previous calculations of inelastic scattering from textured samples, that also find only small differences between them and random powder samples~\cite{Wu1994}.  In contrast, the elastic scattering (i.e., the relative intensities of the Bragg peaks) strongly depends on texture, even when azimuthally averaged, as the elastic scattering for a particular peak corresponds to a distinct  line in reciprocal space, defined by the intersection of the Ewald sphere with the Polanyi sphere, with no large averaging over the surface of the sphere.  As the intensity of the Bragg peaks themselves are so sensitive to texture, and the texture itself changes under shock compression due to plastic flow, we cannot easily extract the DW factor from the relative intensities of the Bragg peaks.

\begin{figure}
     \includegraphics[width=1.0\columnwidth]{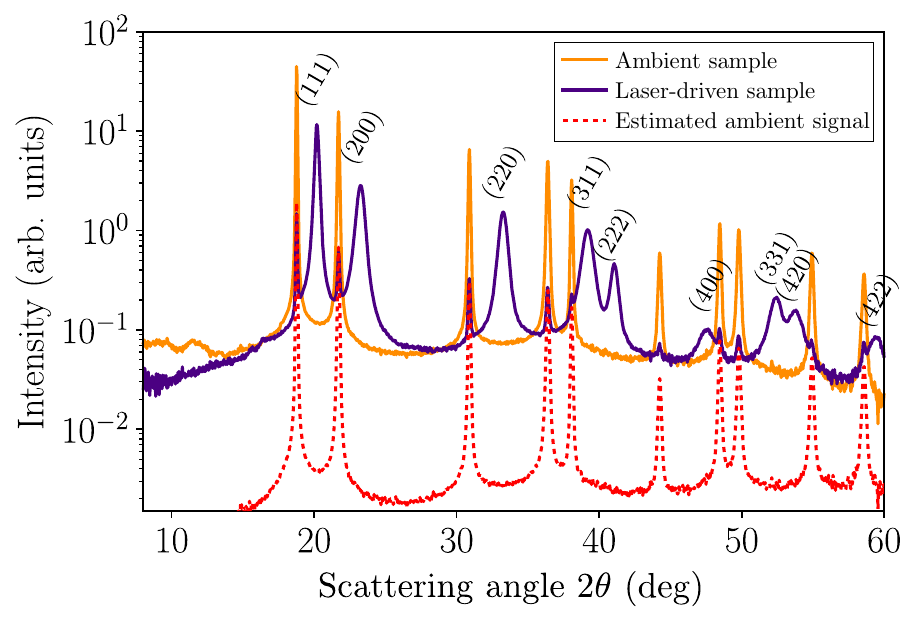}
     \caption {The x-ray diffraction pattern from a sample shock compressed to a relative volume of 0.93 (a pressure, according to the SESAME 3336 EOS, of 12 GPa).  Diffraction from a 25-$\upmu$m thick ambient sample is also shown, as is the calculated signal (taking into account photoelectric absorption) for the unshocked region of a shocked target, such that the thickness of  shocked region is $x=0.96$. }
    \label{F:Shock_thickness}
\end{figure}

Having shown above that the azimuthally averaged TDS is insensitive to texture, we now consider data from shock-compressed targets.  Note that for each target a diffraction pattern was obtained under ambient conditions and subsequently during the passage of the shock.  The relative intensities of the two patterns can be compared by normalizing them to the incoming x-ray flux (via use of the XGM detectors).  

As noted above, in order to observe the shock-induced changes in the DW factor, exhibited by changes in the intensity of the TDS, we require a large measurable fraction of the Cu target to be shocked at the time when the diffraction pattern is recorded.  This fraction is deduced from the intensity of the diffraction from the thin rear layer of the target which has yet to be shocked, as illustrated in \hyperref[F:Shock_thickness]{Fig.~6}, where we show the diffraction signal from a Cu target shock compressed to a relative volume of 0.93, corresponding to a pressure (according to SESAME 3336) of order 12 GPa.  Note that as well as the shift in the Bragg peaks to higher angles due to the shock compression, we are still recording far weaker Bragg diffraction from a layer of ambient material [the layer of thickness $(1-x)L_{\mathrm{Cu}}$ in \hyperref[F:setup]{Fig.~1}.

The fraction $x$ can be ascertained by comparing the intensity of the diffraction from this unshocked layer with that of the target before shock compression. \hyperref[F:Shock_thickness]{Figure~6} also shows this pre-shot data, but reduced in intensity by an amount corresponding to the x-rays first having to pass through 0.96 of the target of the target (which in this particular case is our deduced shocked fraction), such that the intensity of its Bragg peaks aligns with those from the unshocked regions of the driven target. Note that not every one of the Bragg peaks exactly fits this thickness, and this is due to the effect that the target is moved between the pre-shot and main shot, and there are variations in the sampled texture of the foil.  A least-squares fitting must therefore be performed, by which, in this particular case, we find our error in $x$ is $\pm 0.01$. The full procedure for deducing the shock fraction, $x$, along with the error analysis, is outlined in the Supplementary Material.

In order to have sufficient sensitivity to small shock-induced changes in the TDS, we need the vast majority of the target to be in the shocked state: not only will a large unshocked fraction result in a small overall change in the relevant TDS being observed, but it should also be borne in mind that we wish to ascertain the TDS signal of the shocked material approximately midway between the elastic Bragg peaks of the shocked material, to ensure we differentiate between it and the elastic scattering, yet this is also the region between the Bragg peaks where the TDS minimises.  In addition, at higher and higher shock compressions, the Bragg peaks (and nearby TDS) from the unshocked region will start to encroach at the same scattering angles as those at which we are measuring the TDS from the shocked region. It is also the case that the scattering from the shocked material (our signal), is absorbed within the as-yet unshocked material, causing further degradation in our signal if the shocked fraction is not sufficiently large.  The final errors in any single datum thus vary as function of both shock pressure and $x$.  For the range of compressions observed in our experiment, we find that for the error in  our measurements of the DW factor to be dominated by the error due to the XGM (i.e., measurement of scattering intensity), we require greater than 80\% of the target to be in the shocked state.  This will become evident in the data we present below.

\begin{sidewaysfigure*}
\vskip 9.5cm
     \includegraphics{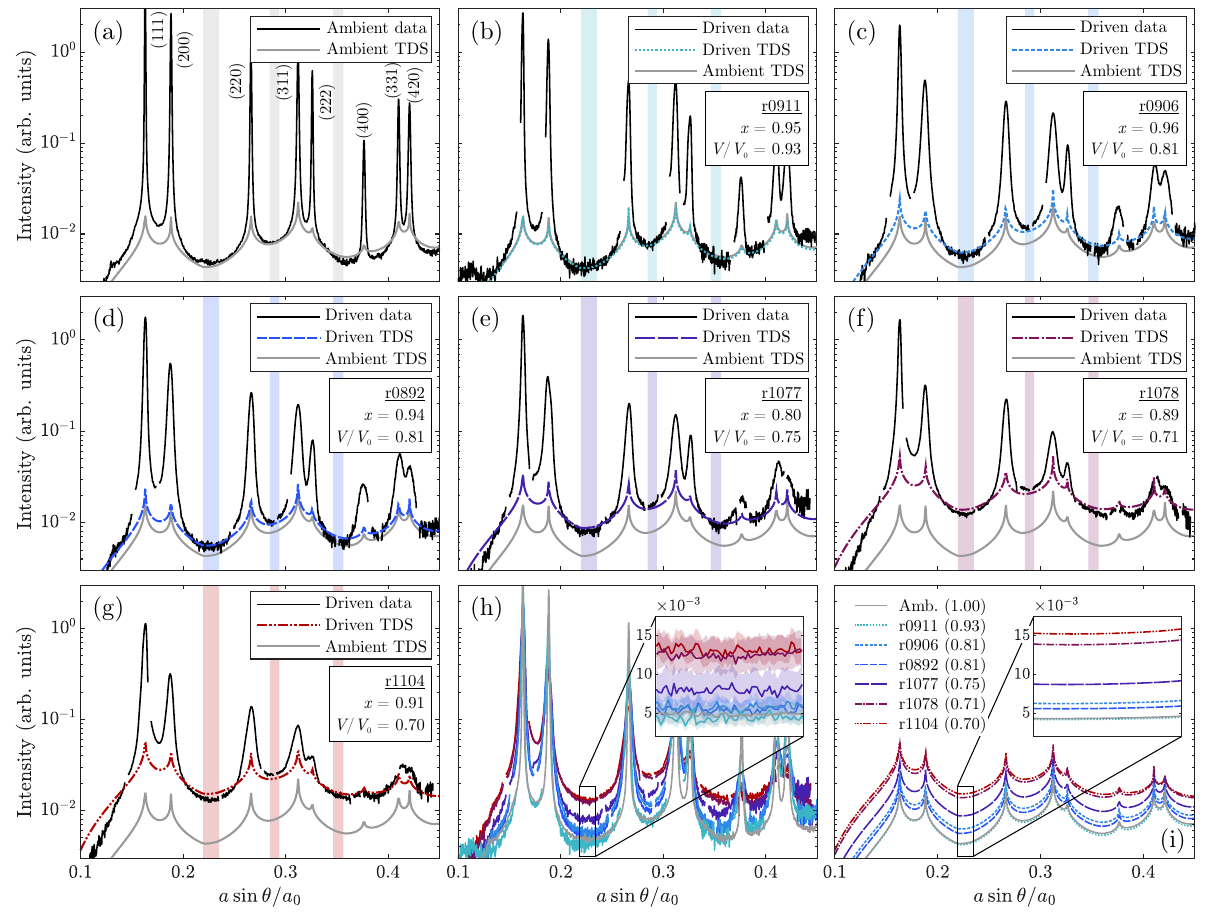}
     \caption{Overview of the exemplary dataset -- including only those shots for which the shock fraction $x\ge0.8$ -- and thermal diffuse scattering (TDS) fitting results. (a-g) Diffracted intensity (solid black lines) and simulated TDS signal (broken colored lines) for compression ratios $V/V_0$ between unity (ambient data) and 0.70. For (b-g), the fitted ambient TDS is shown by a solid gray line for reference. Shaded regions show the domains of the experimental diffraction patterns used to fit Warren's TDS model. (h) Aggregate of the experimental data (labeled by run number and compression ratio) with inset showing variation of the signal within the first fitting window, including $\pm1\sigma$ intervals. (i) Aggregate of simulated TDS signals, labeled similarly. All data are plotted with abscissa $a\sin\theta/a_0$, where $a$ and $a_0$ are the compressed and ambient cubic lattice constants, respectively.}
    \label{F:Warren_data}
\end{sidewaysfigure*}

In \hyperref[F:Warren_data]{Fig.~7(a-g)} we show these diffraction signals for ambient material and for the six data shots that we have for $x>0.8$, where the diffracted intensity is now plotted as a function of $(a \sin \theta/ a_0)$, where $a$ is the lattice spacing of the sample under compression, and $a_0$ the lattice spacing of the ambient material.  On each of the individual plots, we also show the best fit to the TDS for the ambient material, such that the changes in the intensity of the TDS upon compression can be seen for each individual plot.  This effect of shock compression on the magnitude of the diffracted signal can be seen even more clearly when all of the data is plotted together; this is shown in \hyperref[F:Warren_data]{Fig.~7(h)}.  Note, for all of the data shown in \hyperref[F:Warren_data]{Fig.~7} we have removed the low-intensity Bragg peaks from the unshocked material for clarity, and these regions can be seen as breaks in the data at the same scattering angles. In \hyperref[F:Warren_data]{Fig.~7(i)} we show all of the best fits of the TDS scattering for each of the shots: the good agreement between the fitted TDS and the experimental data in \hyperref[F:Warren_data]{Fig.~7(h)} is readily apparent.

It can be seen that there is a systematic change in the intensity of the TDS with shock pressure, and in the regions between the (200) and (220) peaks, the (220) and (311) peaks, and the (222) and (400) peaks, the intensity increases by a factor between two and three at the highest shock pressures, but only starts to rise significantly above a relative volume of 0.81 (a pressure of 52~GPa according to SESAME 3336).  Note also that at high shock compressions the high-order diffraction peaks actually start to become dominated by the TDS, rather than elastic scattering, illustrating the difficulties that would ensue by attempting to  measure the DW factor from the ratios of just the elastic peaks  if the TDS is not taken into account, even if texture were not an issue. Indeed, for the (331) and (420) peaks almost all of the scattering we observe is TDS at a compression of 0.7. The fact that a significant fraction of the intensity of a diffraction peak can actually be due to TDS at high temperatures has long been recognised~\cite{Warren1953,Chipman1959b}.

In order to extract values of the DW factor, we perform a least-squares fit to the data of the predictions of the Warren model for the TDS as a function of $2M$, where we constrain the fit to be in three specific locations in the diffracted signal.  These three positions are midway between the (220)/(220) peaks, the (220)/(311) peaks, and the (222)/(400) peaks, where in each case we fit over a range of angles corresponding to a width of 20\% of the $2 \theta$ separation between the peaks.  These three regions are shown shaded in each of \hyperref[F:Warren_data]{Fig.~7(a-g)}.  These positions are chosen as they correspond to the scattering angles where the TDS significantly dominates over any contribution that could be attributed to the wings of the Bragg peaks.  For example, as can be seen in \hyperref[F:Warren_data]{Fig.~7}, the (311) and (222) peaks are sufficiently close together that the TDS intensity cannot be accurately ascertained.  The best fit for the Warren model for each of the data shots is also shown in \hyperref[F:Warren_data]{Fig.~7(a-g)}, and all of the fits shown together as a function of $(a \sin \theta/ a_0)$ in \hyperref[F:Warren_data]{Fig.~7(i)}.

\begin{figure}
    \includegraphics{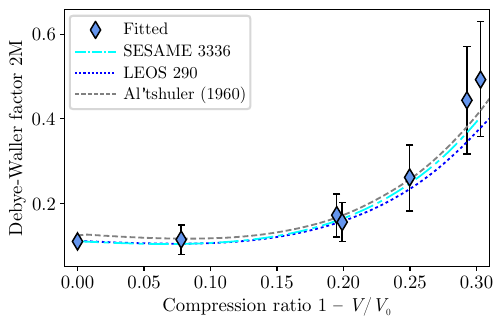}
    
    \caption {The Debye-Waller (DW) factor, $2M$, as a function of compression, sampled at the (pressure-dependent) $2\theta$ angle midway between the (200) and (220) Bragg peaks. Discrete points show the DW factors inferred by fitting the experimental data, using Debye temperatures predicted by LEOS~290. Overlaid are the DW factors calculated using the Hugoniot density, temperature, and Debye temperatures predicted directly by the thermal equations of state SESAME 3336 and LEOS 290. We also show the DW factor calculated using Gr{\"u}neisen-parameter measurements by Al'tshuler~\cite{Altshuler1960}, assuming an ambient Debye temperature of 311~K~\cite{Ho1974}.}
    \label{F:2M_Hugoniot}
\end{figure}

The values of the DW factor as a function of compression corresponding to the fits to the data shown in \hyperref[F:Warren_data]{Fig.~7} for the region midway between the (200) and (220) peaks are shown in \hyperref[F:2M_Hugoniot]{Fig.~8}. Values of the pressure-dependent Debye temperature $\Theta_D$ predicted by LEOS 290 have been used for illustration. It is interesting to note that the DW factor is predicted by the EOS model to initially slightly decrease upon compression, and the data is evidently consistent with this very effect, albeit with an error bar of a magnitude that would prevent us from claiming to have conclusively observed it.  Such a reduction in $2M$ upon weak shock compression has previously been predicted~\cite{Murphy2008}. At low shock pressures the rise in the Debye temperature has a greater influence than the small increase in material temperature, along with the increase in the length of the scattering vector.  This is due to the fact that at low shock strengths the Hugoniot remains close to the isentrope.  By definition, along an isentrope $(T / \Theta_D)$ remains constant, and thus the increase of $\Theta_D$ with compression leads to a decrease in the $T |{\bf G}|^2/ \Theta_D^2$ (where $\bf{G}$ now corresponds to the point in reciprocal space associated with the scattering vector), as long as the effective Gr{\"u}neisen parameter exceeds 2/3~\cite{Murphy2008}.  At higher shock pressures, as the Hugoniot deviates further from the isentrope, and significant shock heating occurs such that the temperature rise dominates over any increase in the square of the Debye temperature, the DW factor increases.

The temperatures that we deduce will depend on our model of the Debye temperature as a function of compression, for which there are a number of predictions which we can employ.  Here we consider two thermal equations of state that have been used to model shock compressed copper, and to which experimental data were also compared for the EXAFS work referred to previously~\cite{Sio2023}, namely the SESAME EOS~3336 and LEOS~290. Both of these model equations of state make specific predictions both for the Debye temperature itself, and for the temperature along the Hugoniot.

\begin{figure*}
    \includegraphics[angle=0,width=\textwidth]{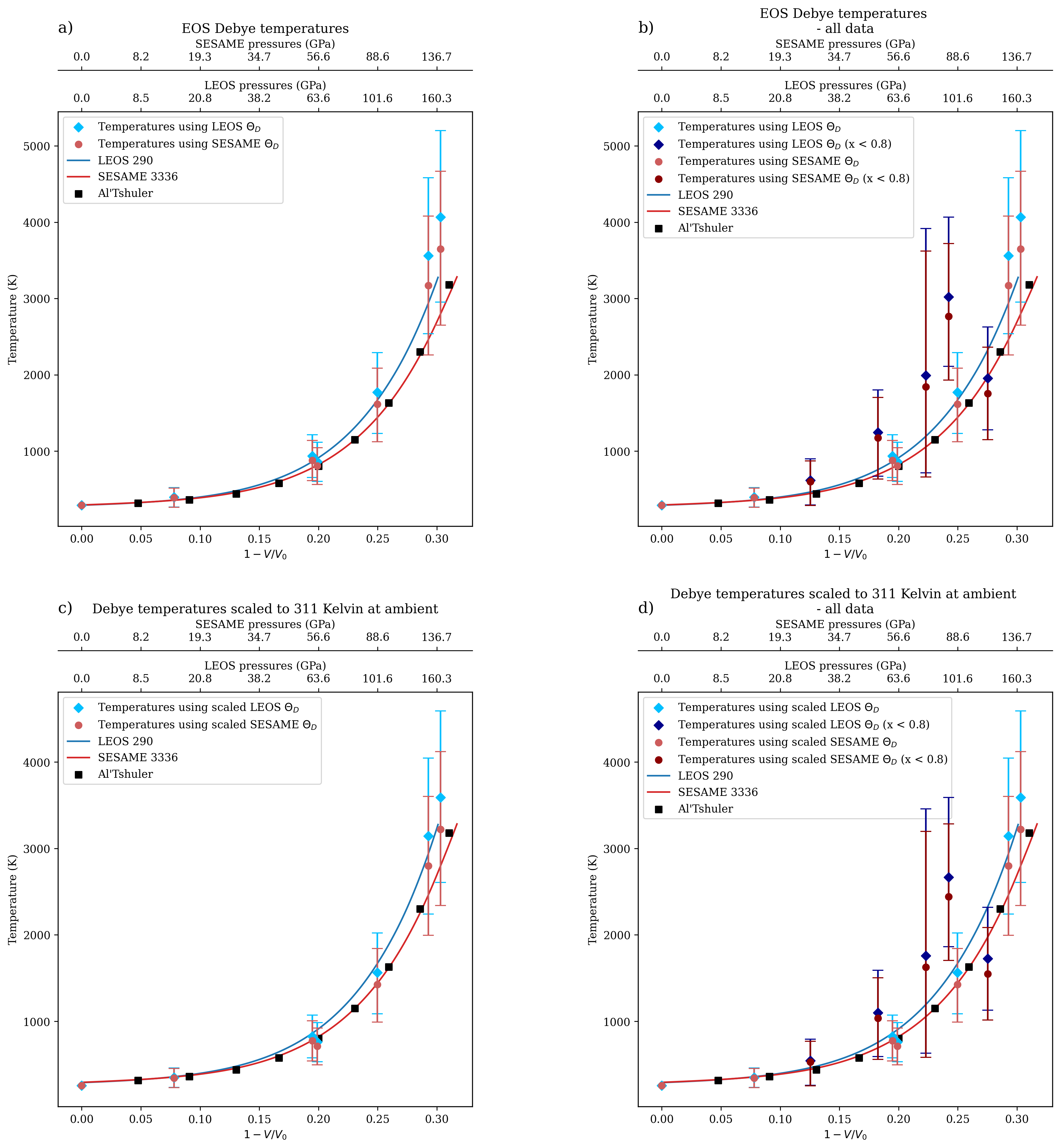}
    \caption {Temperature as a function of compression for: (a) those data shots where $x>0.8$ and using $\Theta_D$ predicted by the equation of state (EOS) itself; (b), as (a), but including data with lower shock fractions; (c), as (a), but assuming an initial $\Theta_D$ of the experimental value of 311~K; and (d), as (c), but including data with lower shock fractions.  In all cases we also show the temperatures predicted by Al'tshuler~\cite{Altshuler1960}.  The upper x-axis shows shock pressures for given compressions predicted by  the SESAME~3336 and LEOS~290 EOS.}
    \label{F:Temperature}
\end{figure*}

In \hyperref[F:Temperature]{Fig.~9}, we plot the temperature as a function of compression that our data imply if we use the Debye temperatures from the SESAME 3336 and LEOS 290 equations of state, alongside the temperatures on the Hugoniot that those models themselves predict.
We also plot the temperatures deduced from some of the first experimental data for shocked Cu obtained at these compressions, due to Al'tshuler and co-workers~\cite{Altshuler1960}, where the temperatures were derived using a Mie-Gr{\"u}neisen model, and represent an extension of lower-pressure data collected in the original work of Walsh {\it et al.}~\cite{Walsh1957}  \hyperref[F:Temperature]{Figure~9(a)} shows the temperatures deduced from the data set shown in \hyperref[F:Warren_data]{Fig.~7} -- i.e., those data for which $x>0.8$. In order to show how the errors in our measurement increase for those shots with lower shock fractions, in \hyperref[F:Temperature]{Fig.~9(b)} we have plotted the temperatures deduced for our full data set, which includes shots for which $x$ can be as low as 0.6.  A comparison of these two plots demonstrates the increase in our errors for lower shock fractions, though as noted above, the way that the errors propagate is a non-trivial function of both shock fraction and the degree of compression, given the way that the scattering from the unshocked and shocked portions of the crystal overlap.

It can be seen that, within the error bars of the experimental data for larger shock fractions, there is very good overall agreement between the temperatures deduced from the TDS and the theoretical predictions, with the Warren model of the data implying that temperatures of order 800~K are achieved under shock compression to a value of $1 - V / V_0$ of 0.2, and rising to over 3000~K when $1 - V / V_0$ reaches 0.3.  It should be noted that both the SESAME~3336 and LEOS~290 models make predictions for the Debye temperature itself, and for Cu at its ambient density and temperature: for both equations of state this value is 331~K.  However, the experimental value for $\Theta_D$ at STP is 311~K~\cite{Ho1974}, and as the temperature we deduce from the DW factor will scale as $\Theta_D^2$, using the experimental value would imply lower temperatures.  Thus we also plot in \hyperref[F:Temperature]{Fig.~9(c,d)} the temperatures we would calculate from these two EOS if we replaced the initial value of $\Theta_D$ by the experimental one, but then used the implied Gr{\"u}neisen parameter as a function of volume to subsequently model $\Theta_D$ under compression.  As can be seen, this leads to slightly lower temperatures in both cases, but the variation is smaller than our experimental error.

We thus conclude that within experimental error the single-shot measurements of the TDS allow us to determine temperatures that are consistent with these EOS models, although we are still reliant on their predictions of the Debye temperature (or Gr{\"u}neisen parameter) to make this claim, much as in the same way that the EXAFS data must rely on the accuracy of potentials within the MD simulations.  Nevertheless, given the importance of being able to make such single-shot temperature measurements in FEL experiments, we believe that the results we have presented here constitute an important step forward in temperature measurements from dynamically compressed solid state matter.

\section{Discussion}\label{S:Discussion}

Whilst the temperatures that we have deduced for shocked copper are in good agreement with EOS models, it is evident from \hyperref[F:Temperature]{Fig.~9} that we are not yet in a position to make  meaningful statements about which EOS predicts the best value for temperature. This is both because of the size of the error bars in our experimental data, as well as a lack of independent measurements of $\Theta_D$ under compression. In this section we discuss how improvements can be made in these areas to the initial data presented here, as well as remarking on other considerations for the applicability of the technique in dynamic compression experiments.

Our main source of error in deducing the DW factor at present is in our measurement of the incident x-ray flux on the target. The x-ray flux before the x-ray focusing lenses is monitored by the XGM, which has a measurement uncertainty of approximately $\pm 10\%$.  IPM diodes, which reside after the the x-ray focusing lenses (and thus are not affected by the lens transmission) provide a more precise reading of the x-ray intensity, but, in this initial experiment, provided signals that were corrupted upon firing of the DiPOLE laser (an issue that did not affect the XGMs).  The data we provide here is thus reliant on calibrating the incident x-ray flux on unshocked targets, with the transmission of the x-ray lenses for the two different x-ray spot sizes used being determined prior to all of the DiPOLE shots, and then using the  XGM readings to determine the incident x-ray flux (as described in the Supplementary Material).  Future improvements would include better shielding of the IPM diodes, such that they provide meaningful results on DiPOLE shots themselves.

Two other factors give rise to errors in our measurements: the unshocked fraction of the target at the time of data collection, and the scattering from the ablator material. As for the first of these, obtaining diffraction patterns where the vast majority of the target is uniformly shocked is clearly an advantage, but should not be an issue in future experiments.  It should be noted that  the results presented here were obtained as a small part of the first user experiment (the EuXFEL 2740 community proposal) to be performed using the DiPOLE laser at the EuXFEL HED instrument, during which several other types of  proof-of-principle studies were undertaken.  As such, a limited number of shots were available.  Furthermore, on this first experiment the vast majority of the data was collected at relatively low repetition rates. In principle DiPOLE can operate at 10~Hz, and, for at least a short duration, it has been demonstrated that DiPOLE can be operated as diffraction data is collected at Hz rates~\cite{Eggert2024}. As a result, in future work, we envisage a large increase in total data collected, and no issues in timing the shock to reach to almost exactly the rear of the target.

The scattering from the ablator layer is also a factor that needs further study.  Firstly, it would be useful to have extensive studies of the scattering from the ablator at different pressures.  For the work here, where Kapton was used, we do not expect any particular structure to form in the regions where we are measuring the TDS from the Cu, and in any case the majority of the scattering from the ablator is due to Compton scattering.  The various competing effects of the elastic and incoherent scattering should be taken into account when evaluating competing ablator materials (e.g., diamond). We also note that the use of any ablator will, to a degree, limit the use of this TDS technique to targets with a high enough atomic number, such that the inelastic TDS from them dominates any scattering from the ablator.  Furthermore, for a fixed photon energy, for lower $Z$ targets Compton scattering will become more of an issue. 

We note that the quantity being measured by recording the TDS (the DW factor) is almost identical to that which is deduced from the EXAFS technique. When applied to the field of dynamic-compression science, each technique will find a range of applicability depending upon the experimental facility and the target under study.  In any event, we are essentially measuring $T / \Theta_D^2$, and thus are reliant on a model of $\Theta_D$ under compression, which is the same as knowing the compression-dependent Gr{\"u}neisen parameter (or as in the case with the recent EXAFS data, comparison is made directly with MD, and thus reliant on the fidelity of the potential used).

It is thus of interest to ask: could we obtain information about $\Theta_D$ itself, thus allowing us to directly infer temperature? We have discussed in the introduction that it has been shown that spectrally resolved IXS from phonons can be obtained on these timescales~\cite{McBride2018,Descamps2020a,Wollenweber2021}, but owing to the high resolution required, these measurements are very photon hungry.  It would be useful to analyze whether using such a technique simply to glean a value for the highest phonon energy in the system (an effective measure of $\Theta_D$), rather than explicitly attempting to directly ascertain temperature via detailed balance, may require fewer shots.

Furthermore, at least in a restricted set of conditions, it may be possible to measure $\Theta_D$ by the TDS technique discussed here.  Within the work we have presented, and in our use of the Warren model, we have assumed the high temperature limit, such that the number of phonons per mode of frequency $\omega$ is simply proportional to $k_{\rm B}T/ \hbar \omega$.  In this case the TDS simply depends on $T / \Theta_D^2$.  However, if the temperature is significantly lower than the Debye temperature, then the amplitude of the higher-energy phonon modes starts to be determined by their zero-point motion, which modifies the form of the TDS such that it also becomes a function of $T / \Theta_D$~\cite{Herbstein1955}, and thus the detailed form of the TDS allows $T$ and $\Theta_D$ to be determined separately.  Of course, in many circumstances, given that typical Debye temperatures are of order room temperature (as here), the high temperature approximation will hold in the majority of cases.  However, the above issue would be interesting to explore in materials with high Debye temperatures (e.g., diamond, although as noted above, Compton scattering may preclude this, at least at the photon energies used here), or if dynamically compressing materials initially at very low temperature, especially if those materials are compressed quasi-isentropically, rather than shock compressed, such that the temperature remains low compared with the Debye temperature along the compression path.  Indeed, in so-called quasi-isentropic compression, the dominant form of heating of the material to temperatures above the Debye temperature will be due to the plastic work performed, which in turn is determined by material strength, itself a material property of great interest at ultra-high strain-rates~\cite{Wark2022}. 

The above considerations lead us then to address further improvements to the modeling. In the work presented here we have used the classic model of Warren to calculate the TDS, and it is evident that in the regions studied, good agreement is found between it and the data.  We choose this model for its simplicity of implementation, which has allowed us to readily adapt it for arbitrary ODFs, showing how the azimuthally integrated TDS is quite insensitive to texture. Several improvements to the Warren model can easily be incorporated in the future.  For example, and related to the low $T$ situation referred to above, it has been shown that it is straightforward to take into account situations where $T$ is no longer of order or higher than $\Theta_D$~\cite{Herbstein1955}.  Also, Warren uses the Debye approximation of a linear dispersion relation for the phonons, but modification of the model to include non-linear dispersion as well as anisotropic materials has been developed~\cite{Walker1972}, and could readily be implemented.
Furthermore, much more sophisticated calculations of the TDS, and predictions of $\Theta_D$ under compression, can be made from Density Functional Theory (DFT), with open-source software such as the package \textsc{ab2tds}~\cite{AB2TDS} which is based on the formalism of Xu {\it et al.}~\cite{XuChiang2005},  available for such a purpose.  Indeed, if we wish to apply this TDS method to more complicated systems such as compounds, then the simple model of Warren will not suffice.  In addition, with more accurate measurements of the TDS, it is likely that anharmonic effects will need to be taken into account, which are not incorporated into the relatively simple analysis presented here~\cite{Maradudin1963}.

Notwithstanding all of the improvements that could evidently be made to these initial measurements, we conclude in noting that we have used the output from an x-ray FEL to measure  intensity of the spectrally integrated but angularly resolved inelastic x-ray TDS from laser-shocked copper foils. Simulations using an adapted version of the classic model of Warren show that the azimuthally averaged TDS signal is insensitive to texture, but strongly dependent upon the DW factor, effectively giving a measure of $T/ \Theta_{D}^2$. Using compression-dependent Debye temperatures from the SESAME~3336 and LEOS~290 EOS, we find temperatures along the Hugoniot that agree well with predicted values.  We believe that in the future the experimental errors in these single-shot measurements could be significantly reduced by more accurate measurements of the incident x-ray flux, and larger data sets with shock fractions, $x$, very close to unity.  We posit that this technique affords a relatively straightforward method to obtain single-shot information on the temperature of a range of dynamically compressed materials on femtosecond timescales.

\section*{Supplementary Material}
See the Supplementary Material for details of the algorithm used to isolate the x-ray scattering signal from the shock-compressed copper alone (including the requisite calculation of the shock fraction $x$), and the overall structure of the model we used to predict both the elastic and inelastic components of the x-ray scattering using the classic theory of Warren.

\begin{acknowledgments}

JSW and PGH gratefully acknowledge support from EPSRC under research grant 	EP/X031624/1. DJP and TS appreciate support from AWE via the Oxford Centre for High Energy Density Science (OxCHEDS).

We acknowledge the European XFEL in Schenefeld, Germany, for provision of X-ray free electron laser beam time at the Scientific Instrument HED (High Energy Density Science) and would like to thank the staff for their assistance. The authors are indebted to the Helmholtz International Beamline for Extreme Fields (HIBEF) user consortium for the provision of instrumentation and staff that enabled this experiment. The data is available upon reasonable request [doi: 10.22003/XFEL.EU-DATA-002740-00].

We acknowledge DESY (Hamburg, Germany), a member of the Helmholtz Association HGF, for the provision of experimental facilities. Parts of this research were carried out at PETRA III (beamline P02.2).

Part of this work was performed under the auspices of the U.S.\ Department of Energy by Lawrence Livermore National Laboratory under Contract No.\ DE-AC52-07NA27344 and was  supported by the Laboratory Directed Research and Development Program at LLNL (Project No.\ 21-ERD-032).

Part of this work was performed under the auspices of the U.S.\ Department of Energy through the Los Alamos National Laboratory, operated by Triad National Security, LLC, for the National Nuclear Security Administration (Contract No.\ 89233218CNA000001).

Research presented in this article was supported by the Department of Energy, Laboratory Directed Research and Development program at Los Alamos National Laboratory under Project No.\ 20190643DR and at SLAC National Accelerator Laboratory, under Contract No.\ DE-AC02-76SF00515.

This work was supported by Grants No.\ EP/S022155/1 (MIM, MJD) EP/S023585/1 (AH, LA) and EP/S025065/1 (JSW) from the UK Engineering and Physical Sciences Research Council. JDM is grateful to AWE for the award of CASE Studentship P030463429.

EEM and AD were supported by the UK Research and Innovation Future Leaders Fellowship (MR/W008211/1) awarded to EEM. 

DE and DS from Univ.\ de Valencia thank the financial support by the Spanish Ministerio de Ciencia e Innovación (MICINN) and the Agencia Estatal de Investigación (MCIN/AEI/10.13039/501100011033) under grants PGC2021-125518NB-I00 and PID2022-138076NB-C41 (cofinanced by EU FEDER funds), and by the Generalitat Valenciana under grants CIPROM/2021/075, CIAICO/2021/241 and MFA/2022/007 (funded by Next Generation EU PRTR-C17.I1).

NJH and AG were supported by the DOE Office of Science, Fusion Energy Science under FWP 100182. This material is based upon work supported by the Department of Energy National Nuclear Security Administration under Award Number DE-NA0003856.

YL is grateful for the support from the Leader Researcher program (NRF-2018R1A3B1052042) of the Korean Ministry of Science and ICT (MSIT).

KA, KB, ZK, HPL, RR and TT thank the DFG for support within the Research Unit FOR 2440. 

BM and RSM acknowledge funding from the European Research Council (ERC) under the European Union’s Horizon 2020 research and innovation programme (Grant agreement No.\ 101002868).

GWC and T-AS recognize support from NSF Physics Frontier Center Award No.\ PHY-2020249 and support by the U.S.\ Department of Energy National Nuclear Security Administration under Award No.\ DE-NA0003856, the University of Rochester, and the New York State
Energy Research and Development Authority.

SM, HG, and JC are funded by the European Union (ERC, HotCores, Grant No.\ 101054994). Views and opinions expressed are however those of the author(s) only and do not necessarily reflect those of the European Union or
the European Research Council. Neither the European Union nor the granting authority can be held responsible for them. 

The work of DK, DR, JR, and MS was supported by Deutsche Forschungsgemeinschaft (DFG—German Research Foundation) Project No.\ 505630685.

SP acknowledges support from the GotoXFEL 2023 AAP from CNRS.

\end{acknowledgments}

\section*{Author declarations}
The authors have no conflicts or competing interests to disclose.

\section*{Data availability}
The x-ray diffraction data obtained from the experiment at EuXFEL described in this article are available at \href{https://doi.org/10.22003/XFEL.EU-DATA-002740-00}{doi:10.22003/XFEL.EU-DATA-002740-00}.

\section*{References}
%

\end{document}


\title{Supplementary material: Femtosecond temperature measurements of laser-shocked copper deduced from the intensity of the x-ray thermal diffuse scattering}

\author{J.S. Wark~\orcidlink{0000-0003-3055-3223}}\email{justin.wark@physics.ox.ac.uk}
\affiliation{Department of Physics, Clarendon Laboratory, University of Oxford, Parks Road, Oxford OX1 3PU, UK\looseness=-1}

\author{D.J. Peake \orcidlink{0000-0002-5992-6954}}
\affiliation{Department of Physics, Clarendon Laboratory, University of Oxford, Parks Road, Oxford OX1 3PU, UK\looseness=-1}

\author{T. Stevens~\orcidlink{0009-0006-8355-3509}}
\affiliation{Department of Physics, Clarendon Laboratory, University of Oxford, Parks Road, Oxford OX1 3PU, UK\looseness=-1}

\author{P.G. Heighway~\orcidlink{0000-0001-6221-0650}}\email{patrick.heighway@physics.ox.ac.uk}
\affiliation{Department of Physics, Clarendon Laboratory, University of Oxford, Parks Road, Oxford OX1 3PU, UK\looseness=-1}

\author{Y. Ping~\orcidlink{0000-0002-4879-9072}}
\affiliation{Lawrence Livermore National Laboratory, Livermore, CA 94550, USA\looseness=-1}

\author{P. Sterne~\orcidlink{0000-0002-6398-3185}}
\affiliation{Lawrence Livermore National Laboratory, Livermore, CA 94550, USA\looseness=-1}


\author{B. Albertazzi}\affiliation{Ecole Polytechnique, Palaiseau, Laboratoire pour l'utilisation des lasers intenses (LULI), CNRS UMR 7605 Route de Saclay 91128 PALAISEAU Cedex, France\looseness=-1}

\author{S.J. Ali~\orcidlink{0000-0003-1823-3788}}\affiliation{Lawrence Livermore National Laboratory, Livermore, CA 94550, USA\looseness=-1}

\author{L. Antonelli~\orcidlink{0000-0003-0694-948X}}\affiliation{University of York, School of Physics, Engineering and Technology, Heslington York YO10 5DD, UK\looseness=-1}

\author{M.R. Armstrong~\orcidlink{0000-0003-2375-1491}}\affiliation{Lawrence Livermore National Laboratory, Livermore, CA 94550, USA\looseness=-1}

% ...

\date{\today}

\begin{abstract}

This Supplementary Material details the algorithm used to isolate the x-ray scattering signal from the shock-compressed copper alone (including the requisite calculation of the shock fraction $x$), and the overall structure of the model we used to predict both the elastic and inelastic components of the x-ray scattering using the classic theory of Warren.

\end{abstract}

\maketitle

\section{Calculating the shock mass-fraction and isolating the compressed copper signal}

\subsection{Introduction}
In general, the diffuse signal from our layered targets contains contributions from both ambient and compressed copper, as well as the sacrificial Kapton-B ablator. In addition, x-rays passing through the copper regions of the sample are attenuated with distance traveled (primarily by photoelectric absorption), giving the scattering contributions of successive layers of material an angular-dependent weight. To isolate the structure factor of the compressed copper \textit{alone} -- and thus determine its temperature -- it is necessary to carefully subtract all other contributions.

Subtracting the contribution from the remaining ambient (i.e., uncompressed) copper towards the rear surface of the target requires knowledge of the mass fraction of shocked material, $x$. To calculate $x$, we compared the intensities of the ambient copper peaks in the main-shot data with the corresponding intensities in the pre-shot (undriven) data: the greater the fraction of shocked material, the lower the intensity of the ambient peaks. As well as the reduction in ambient material during the shock transit, x-ray attenuation from both the compressed and ambient layers in the sample also diminishes the ambient intensity. Moreover, the intensity of the pattern is also affected by the region of the detector on which the signal is detected. Such effects make determination of the shock-fraction $x$ nontrivial.

In this section, we will detail the algorithm that incorporates the intrinsic and extrinsic corrections to the total measured diffraction signal, deduces the shock mass fraction $x$, and isolates the compressed copper scattering.

\subsection{Data collection}
The intensity data collected across the pair of 2D Varex x-ray detectors is regrouped into equally sized bins in $(2\theta, \phi)$-coordinates using the \textit{pyFAI} Python library, with the x-ray detector geometry having been refined using the Debye rings from powderlike CeO\textsubscript{2} calibrants. Azimuthal integration is performed by calculating the ratio of the intensity-weighted sum and the sum of all pixels in each bin $[2\theta, 2\theta+\Delta]$, such that 
\begin{equation}
    I(2\theta) = \frac{\sum_{2\theta'\in [2\theta, 2\theta+\Delta]}I(2\theta')}{\sum_{[2\theta, 2\theta+\Delta]}}\ . 
\end{equation}

Each detector position and orientation is calculated separately, and combined using the \textit{MultiGeometry} module in the \textit{pyFAI} library.

\subsection{The data: a sum of its constituent parts}
The total per-pixel intensity measured on the detectors can be decomposed into a sum of individual contributions from the Kapton-B ablator, the shocked copper, and remnant unshocked copper:
\begin{equation}
    I(2\theta,\phi) = I_{\textrm{BK}}(2\theta,\phi) + I_{\textrm{Cu}}(2\theta,\phi) + I_{\textrm{Cu,0}}(2\theta,\phi)\ .
    \label{E:superposition}
\end{equation}

For scattering through a small solid angle $\Delta\Omega$, each contribution can be expressed as $I_j = I_0 \times \frac{\Delta \Omega}{\Delta A} (\frac{d\sigma}{d\Omega})_j $ $(j=\textrm{Cu, BK})$, with $I_0$ the incident x-ray flux and $\Delta A$ the effective cross-sectional area of the scattering atoms.

As demonstrated in the seminal work of Hartree and Waller\cite{HartreeWaller1929}, these scattered signals contain both coherent and incoherent (Compton) contributions; the former describes x-rays scattered to the same initial energy, while the latter describes scattering to different energies.
We therefore further decompose each cross-section into its coherent and incoherent components:
\begin{equation}
    \biggr(\frac{d\sigma}{d\Omega}\biggl)_j = S_j\biggl(\frac{d\sigma}{d\Omega}\biggr)_{\textrm{T}}+\ \tilde{s}_j\biggr(\frac{d\sigma}{d\Omega}\biggl)_{\textrm{KN}}\ .
    \label{E:CohIncoh_decomp}
\end{equation}
The first, coherent scattering term scales with the structure factor $S_j$ and the canonical Thomson cross-section for horizontally polarized radiation,
\begin{equation}
    \biggr(\frac{d\sigma}{d\Omega}\biggl)_{\textrm{T}}=r_e^2 \underbrace{\left(\cos^2 2\theta\cos^2\phi + \sin^2\phi \right)}_{f_{\rm P}}\ ,
    \label{E: Thomsoneq}
\end{equation}
where $r_e$ is the classical electron radius and the bracketed term $f_{\rm P}$ will later be referred to as the polarization factor.

The second, incoherent scattering term scales with the incoherent scattering factor $\tilde{s}_j$ and the generalized cross-section expressed by the Klein-Nishina (KN) equation. In general, radiation of wavelength $\lambda$ scattered from an isolated electron results in the relativistic transfer of momentum $\Delta\lambda$ given by the Compton formula $\Delta\lambda \equiv \lambda'-\lambda= \frac{h}{m_e c} [1-\cos(2\theta)]$. By following the full quantum electrodynamic treatment given by Heitler~\cite{WHeitler1954} (making the approximation of a free, stationary electron), the scattering due to this process is described by the KN formula:
\begin{equation}
   \biggr(\frac{d\sigma}{d\Omega}\biggl)_{\textrm{KN}} = \frac{1}{4}r_e^2\biggr(\frac{\lambda}{\lambda'}\biggl)^2\biggr[ \frac{\lambda}{\lambda'} + \frac{\lambda'}{\lambda} - 2  +4(\mathbf{e}_0\cdot\mathbf{e})^2\biggl]
   \label{E:KNequation}
\end{equation}
where the initial and final polarization vectors satisfy $ (\mathbf{e}_0\cdot\mathbf{e})^2 = \cos^2 2\theta\cos^2\phi + \sin^2\phi$. In the non-relativistic limit $\lambda\approx\lambda'$, the KN cross-section reduces to the classical Thomson scattering formula. For our purposes, the photon energy of $18\textrm{~keV} <<m_ec^2$ means that the Compton effect causes a deviation from the Thomson scattering by only $\sim 3\%$ at high $2\theta$; as such, to good approximation we can equate both the Thomson and Klein-Gordon cross-sections with a single global polarization factor $f_{\rm P}$. 
\\

To treat the bound electrons present in a real system of condensed matter (such as our copper targets), Eqs.~\hyperref[E:KNequation]{(S4)} and \hyperref[E: Thomsoneq]{(S5)} for scattering from free electrons can be modified by the factors derived by Hartree and Waller \cite{HartreeWaller1929}: the (squared) atomic form factor $f^2(Q|Z)$ (discussed in Sec.~\hyperref[ssec:atomic form factor]{S2B}, contained within $S_j(2\theta,\phi)$) for the coherent scattering, and the incoherent scattering factor $\tilde{s}_j(2\theta,\phi)$. These latter factors can be found by considering the amplitude, $\mathcal{A}_{mn}(\mathbf{k}',\mathbf{k})$, of the transition between the many-electron states labeled $n,m$ and an x-ray photon with momentum transfer $\mathbf{Q} \equiv \mathbf{k}'-\mathbf{k}$ .  Again, neglecting photon energy transfer, for a $Z-$electron atom this is given by 
\begin{equation}
\begin{split}
    &\mathcal{A}_{mn}(\mathbf{k}',\mathbf{k}) = \int \Psi^{*}_m \Psi_n \sum_{j=1}^Z e^{i(\mathbf{k}'-\mathbf{k})\cdot\mathbf{r_{\textit{j}}}}\prod_{l=1}^Z d^3 \mathbf{r_{\textit{l}}}\ 
\end{split}
\label{E:A_mn}
\end{equation}
where $d^3 \mathbf{r_{\textit{l}}}$ is the volume element for the $l^{th}$ electron. 
The many-body wave functions $\Psi_n$ can be expressed as symmetrized combinations of single-electron wave functions $\psi^{(j)}_{\alpha}(\mathbf{r_{\textit{j}}}) \ \ (j = 1, ..., Z)$. Expanding Eq.~\hyperref[E:A_mn]{(S6)} in terms of $\psi^{(j)}_{\alpha}(\mathbf{r_{\textit{j}}})$, it can be shown that 
\begin{equation}
    \sum_m\lvert \mathcal{A}_{mn}(\mathbf{k}',\mathbf{k}) \rvert^2 \approx \underbrace{Z - \sum_{\alpha=1}^Z \lvert f_{\alpha\alpha}\rvert^2}_{\tilde{s}} + \underbrace{\lvert \sum_{\alpha=1}^Z f_{\alpha\alpha}\rvert^2}_ {f^2(Q|Z)}
    \label{E:Amn expansion}
\end{equation}
where we have made use of the individual-electron transition amplitude
\begin{equation}
    f_{\alpha\beta} = \int d^3 \mathbf{r_{\textit{l}}}\,\psi^{(l)}_{\alpha} \psi^{(l)}_{\beta} e^{i\mathbf{Q}\cdot\mathbf{r_{\textit{k}}}}\ .
\end{equation}

In the non-relativistic limit,  assuming that the total differential cross-section is given by $ \bigr(\frac{d\sigma}{d\Omega}\bigl) = \bigr(\frac{d\sigma}{d\Omega}\bigl)_T \sum_m\lvert \mathcal{A}_{mn}(\vec{k}',\vec{k}) \rvert^2$, the decomposition in Eq.~\hyperref[E:CohIncoh_decomp]{(S3)} falls out naturally; the third term manifestly describes coherent scattering (since the  atomic form factor $f(Q|Z)\equiv \sum_{\alpha=1}^{Z} f_{\alpha\alpha}$), whilst the previous terms describe radiation scattering from each electron wave-function incoherently, giving $\tilde{s} \approx Z - \sum_{\alpha=1}^Z \lvert f_{\alpha\alpha}\rvert^2$ .

In the expansion in Eq.~\hyperref[E:Amn expansion]{(S7)}, we have assumed that all transitions  $\mathcal{A}_{mn}$ are possible; in reality, certain transitions are forbidden, producing additional terms $f_{\alpha\beta}$ with $\alpha \neq \beta$ in Eq.~\hyperref[E:Amn expansion]{(S7)} and $\tilde{s}$.
\\ 

In addition to the polarization factor $f_{\rm P}$, there exist global correction factors due to the solid angle subtended by each detector pixel ($f_\Omega$) and the composite detector-dependent factor ($f_{\rm D}$), which accounts for both the nonuniform response of the pixel arrays and absorption by the aluminum filter separating them from the vacuum of the target chamber. In addition, there are non-global corrections due to photoelectric absorption through the surrounding layers ($F_{\rm A}$) and through the diffracting layer in question ($F_{\rm SA}$). In full, the total detector signal is given by the master equation
\begin{equation}
\begin{split}
    I ={}&I_0 \times f_{\rm P} f_\Omega f_{\rm D} \\ &\times \biggl[ S^{\textrm{tot}}_{\textrm{BK}} F_{\rm SA}(L_{\textrm{BK}},\mu_\textrm{BK}) e^{-\frac{\mu_{\textrm{Cu}}L_{\textrm{Cu}}}{\cos\zeta}} \\
    &+ S^{\textrm{tot}}_{\textrm{Cu}} e^{-\frac{\mu_{\text{BK}}L_{\textrm{BK}}}{\cos\omega}} F_{\rm SA}(xL_{\textrm{Cu}},\mu_{\textrm{Cu}}) e^{-\frac{\mu_{\text{Cu}}(1-x)L_{\textrm{Cu}}}{\cos\zeta}} \\
    &+ S^{\textrm{tot}}_{\textrm{Cu,0}} e^{-\frac{\mu_{\text{BK}}L_{\textrm{BK}}}{\cos\omega}} e^{-\frac{\mu_{\text{Cu}} xL_{\textrm{Cu}}}{\cos\omega}} F_{\rm SA}((1-x)L_{\textrm{Cu}},\mu_{\textrm{Cu}})\biggr] \\ 
\end{split}
\label{E:master_eq}
\end{equation}
where $S^{\textrm{tot}}_j \equiv S_j + \tilde{s}_j$. For brevity, we have neglected to repeatedly write the arguments $(2\theta,\phi)$ in any of the above functions, with the understanding that, in fact, every function varies with scattering direction; we only include variables that parametrize the functions.

The exponential factors in Eq.~\hyperref[E:master_eq]{(S9)} represent the absorption of the x-rays by layers other than the layer in question, while the self-attenuation factors give the absorption of the layer itself. For example, the scattering contribution from the compressed copper -- denoted by $I_{\textrm{Cu}}$ in Eq.~\hyperref[E:superposition]{(S2)} -- is first attenuated by the factor
\begin{equation*}
    F_{\rm A}(L_{\textrm{BK}},\mu_{\textrm{BK}}) = e^{-\frac{\mu_{\text{BK}}L_{\textrm{BK}}}{\cos\omega}}
\end{equation*}
due to the incident beam having to traverse a Lagrangian thickness $L_{\textrm{BK}}$ of Kapton with linear attenuation coefficient $\mu_{\textrm{BK}}$ at angle $\omega$ to the shock direction, before being attenuated further by the compressed copper itself by factor $F_{\rm SA}(xL_{\textrm{Cu}},\mu_{\textrm{Cu}})$ due to the beam traversing a Lagrangian thickness $xL_{\textrm{Cu}}$ of copper with attenuation coefficient $\mu_{\textrm{Cu}}$, and finally suffering absorption by a further factor
\begin{equation*}
    F_{\rm A}((1-x)L_{\textrm{Cu}},\mu_{\textrm{Cu}}) = e^{-\frac{\mu_{\text{Cu}}(1-x)L_{\textrm{Cu}}}{\cos\zeta}}
\end{equation*}
due to the scattered beam having to travel through a Lagrangian thickness of $(1-x)L_{\textrm{Cu}}$ at emergent angle $\zeta$. In the above, $x$ denotes the mass fraction of copper that has been consumed by the shock wave at the instant at which the target is illuminated by the x-rays. The absorption factors applied to each layer's contribution will be elaborated upon further in Sec.~\hyperref[ssec:attenuation]{S1D}.

The purpose of the algorithm we will describe shortly is to isolate the structure factor of the compressed copper only. That is, of the many terms in Eq.~\hyperref[E:master_eq]{(S9)}, we seek to isolate (and subsequently model) $S_{\textrm{Cu}}(2\theta,\phi)$. To do so, every remaining term in Eq.~\hyperref[E:master_eq]{(S9)} must either be measured separately or modeled. Here, we summarize how each term was dealt with in turn.

\begin{figure*}
    \includegraphics{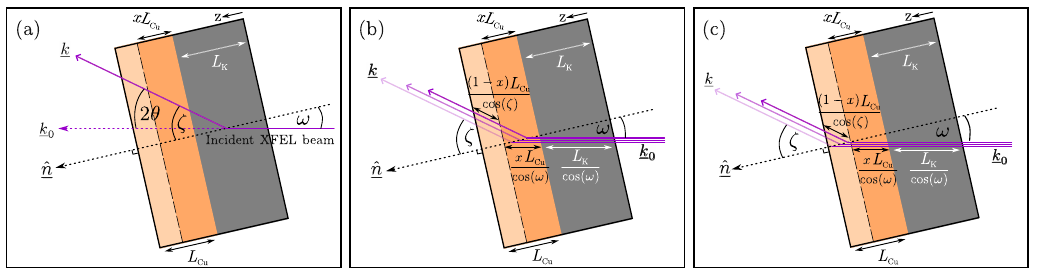} 
    \caption{(a) X-rays scattered by the Kapton-B layer are attenuated by the copper layers. (b) X-rays scattered by the compressed copper region are attenuated by the rear surface ambient layer, as well as within the compressed region (self-attenuation). (c) X-rays scattered by the remnant ambient layer are attenuated by the preceding compressed copper layer, as well as experiencing self-attenuation within the ambient layer.}
    \label{F:scattering_geometry_attenuations_all}
\end{figure*}

The incident beam intensity $I_0$ was ascertained for each shot with an x-ray gas monitor (XGM). This factor was typically measurable to a precision of $\pm10\%$, and is in fact the dominant source of statistical error in our analysis; a full discussion will be given in Sec.~\hyperref[ssec:photonflux]{S1F}.

The polarization factor $f_{\rm P}$ was derived from the Thomson scattering differential cross-section for a horizontally polarized beam:
\begin{equation}
    f_{\rm P}(2\theta,\phi) = \cos^2 2\theta\cos^2 \phi + \sin^2\phi\ ,
\end{equation}
where the `horizontal' plane is spanned by the $\phi=0^\circ$ and $\phi=180^\circ$ half-planes.

The solid angle subtended by the detector pixels with unit normal $\hat{\mathbf{n}}$ at displacement $\mathbf{r}$ from the sample--x-ray interaction point was taken to be
\begin{equation}
    f_\Omega(2\theta,\phi) = \left(\frac{p}{r}\right)^2 \left(\hat{\mathbf{r}}\cdot\hat{\mathbf{n}}\right)\ ,
\end{equation}
where, for the Varex detectors, the pixel size $p$ is 150~$\upmu$m.

The detector-dependent corrective factor $f_{\rm D}(2\theta,\phi)$, which accounts for both nonuniform gain over the detector arrays and absorption by the nonplanar aluminum filters covering them, was determined by calculating the pixel-by-pixel correction needed to ensure that the fluorescence signal from an yttrium-aluminium-garnet (YAG) single crystal was, after factoring out polarization and solid-angle corrections, completely uniform. Details of the flat-fielding procedure are available in Ref.~\onlinecite{Gorman2024}.

The structure factors of the Kapton-B ablator, $S_{\textrm{BK}}$, and the remnant uncompressed copper layer, $S_{\textrm{Cu,0}}$, were measured directly using isolated Kapton targets and undriven Kapton/Cu targets, respectively. More details are provided in Secs.~\hyperref[ssec:kaptonB]{S1E} and \hyperref[ssec:method]{S1G}.

As alluded to above, the various attenuation effects were modeled directly using the known metrology of the targets, the known x-ray attenuation coefficients at 18~keV, and the shock fraction $x$. The model will be described next in Sec.~\hyperref[ssec:attenuation]{S1D}.

Finally, we calculate the shock fraction $x$ using an x-ray-based method that compares the relative intensities of the diffraction peaks from the compressed and remnant uncompressed fractions of the target; the method is given in Sec.~\hyperref[ssec:attenuation]{S1D}.

\subsection{\label{ssec:attenuation} Attenuation corrections and calculation of shock fraction}

We modeled all x-ray attenuation effects using the Beer-Lambert law, according to which x-rays passing through an infinitesimal (Eulerian) thickness $dz$ are reduced in intensity $I$ by an amount $dI=-I \mu dz$, where $\mu$ is the linear attenuation coefficient at a given mass density $\rho$ and photon energy $E$. Note that under the conditions of uniaxial compression by factor $v>1$, the linear attenuation coefficient (which is directly proportional to $\rho$) scales with $v$, while the Eulerian path length decreases by the same factor. In other words, the total mass traversed by the beam is invariant. For this reason, the product $\mu dz$ is constant, and we can therefore use ambient-density linear attenuation coefficients and ambient (Lagrangian) thicknesses throughout.

Given the asymmetric scattering geometry with x-rays incident at $\omega=22.5^\circ$ to the target normal, x-rays scattered by the same $2\theta$ angle can travel through different thicknesses of material depending on their azimuthal angle of emergence $\phi$. It is therefore useful to define the scattering angle $\zeta$ between the target's unit normal $\hat{\mathbf{n}}$, and diffracted wavevector $\mathbf{k}$ (as shown in \hyperref[F:scattering_geometry_attenuations_all]{Fig.~S1}) which satisfies
\begin{align}
    \cos\zeta &= \frac{\hat{\mathbf{n}}\cdot\mathbf{k}}{|\mathbf{k}|} \\
    &= \cos\omega\cos2\theta - \sin\omega\sin2\theta\cos\phi\ .
\end{align}
These equations allow us to map the escape angle $\zeta$ over the face of each detector.

The total attenuation suffered by the contribution of each scattering layer is a combination of absorption from surrounding layers and absorption within the layer itself. To simplify our analysis, we first make the reasonable approximation that x-ray absorption within the Kapton-B ablator layer is negligible: the x-ray energy used (18~keV) is much greater than the K-edge of carbon ($<1$~keV), meaning $\mu_{\textrm{BK}}$ is essentially zero for our purposes. By contrast, even at these hard--x-ray energies, the attenuation length in copper is comparable to the layer thickness ($\mu_{\rm Cu}^{-1}$ = 25~$\upmu$m), meaning absorption cannot be neglected.

To calculate the attenuation owed to `other' layers, we simply integrate the Beer-Lambert law directly. The intensity $I$ of an undeflected x-ray beam with an initial intensity $I(0)$, having traversed a depth $z$ of a uniform region of the sample at an angle $\alpha$ to the sample normal $\hat{\mathbf{n}}$, is
\begin{align}
    I(z) &= I(0) \exp\left(-\frac{\mu z}{\cos\alpha}\right) \\
    &= I(0)F_{\rm A}(\alpha|z,\mu)\ .
\end{align}
To derive the \emph{self}-attenuation of a scattered x-ray beam from a given layer, one can simply divide the scattering layer into sub-layers of infinitesimal thicknesses $dz$. Each sub-layer is treated as an independent scattering region with x-rays incident at an angle $\omega$, and scattered to an angle $\zeta$ to the sample normal; for each sub-layer, the total attenuation is given by $ e^{-\mu z/\cos\omega}e^{-\mu (L-z)/\cos\zeta}dz$. Integrating over the width of the layer $L$ gives the self-attenuation factor:
\begin{equation}
\begin{split}
F_{\rm SA}&(\zeta |\omega,L,\mu) = \\ &\frac{1}{\mu}\left[\frac{1}{\cos\zeta} - \frac{1}{\cos\omega} \right] ^{-1} \left( e^{-\frac{\mu L}{\cos\omega}} - e^{-\frac{\mu L}{\cos\zeta}}\right)\ .
\end{split}
\end{equation}
Note that, defined thus, the self-attenuation factor is \emph{extensive}, i.e., it scales with the layer thickness:
\begin{equation*}
    \lim_{\mu\to0} F_{\rm SA}(\zeta|\omega,L,\mu) = L\ .
\end{equation*}

Neglecting absorption from the Kapton layer, the combination of attenuation factors for the compressed copper layer (of Lagrangian thickness $xL_{\textrm{Cu}}$) is
\begin{equation}
    F_{\textrm{Cu}}(x) = F_{\rm SA}(\zeta|\omega,xL_{\textrm{Cu}},\mu_{\textrm{Cu}}) F_{\rm A}(\zeta|(1-x)L_{\textrm{Cu}},\mu_{\textrm{Cu}})\ ,
\end{equation}
while that of the rear-surface ambient copper layer [of Lagrangian thickness $(1-x)L_{\textrm{Cu}}$] is
\begin{equation}
    F_{\textrm{Cu,0}}(x) = F_{\rm A}(\omega|xL_{\textrm{Cu}},\mu_{\textrm{Cu}}) F_{\rm SA}(\zeta|\omega,(1-x)L_{\textrm{Cu}},\mu_{\textrm{Cu}})\ .
\end{equation}
Since $ \lim_{\mu\to 0} F_{\rm SA}(\zeta |\omega,L,\mu) =L$, the attenuation factor for the Kapton ablator reduces to
\begin{equation}
    F_{\textrm{BK}} = L_{\textrm{BK}}F_{\rm A}(\zeta|L_{\textrm{Cu}},\mu_{\textrm{Cu}})\ .
\end{equation}
For all production samples described in the Main Article, $L_{\textrm{BK}}$ = 50~$\upmu$m and $L_{\textrm{Cu}}$ = 25~$\upmu$m.

It is through the expression for the ambient signal attenuation that we estimate the proportion of shocked copper, $x$; after isolating the ambient structure factor $S_{\textrm{Cu,0}}$ from the pre-shot (undriven) data, we find a value of $x$ which minimizes the following loss function, a sum conducted over the ambient Bragg peaks:
\begin{equation}
     \mathcal{L}(x) = \sum_{hkl} \biggr|I_{hkl}^{(\rm shot)} -  f_{\textrm{tot}} F_{\rm Cu,0}(x) S_{\textrm{Cu,0}}\biggl|
\end{equation}
where $f_{\textrm{tot}} = f_{\rm P}f_{\Omega}f_{\rm D}$. That is, we find $x$ such that the modeled Bragg peak intensities from a layer of ambient copper of thickness $(1-x)L_{\textrm{Cu}}$ -- appropriately attenuated by $xL_{\textrm{Cu}}$ of compressed copper and its own self-attenuation -- are, on average, closest to the same ambient peak intensities measured on the driven shot.

\subsection{\label{ssec:kaptonB} Kapton-B subtraction} 
\begin{figure}[b]
\includegraphics[width=0.95\columnwidth]{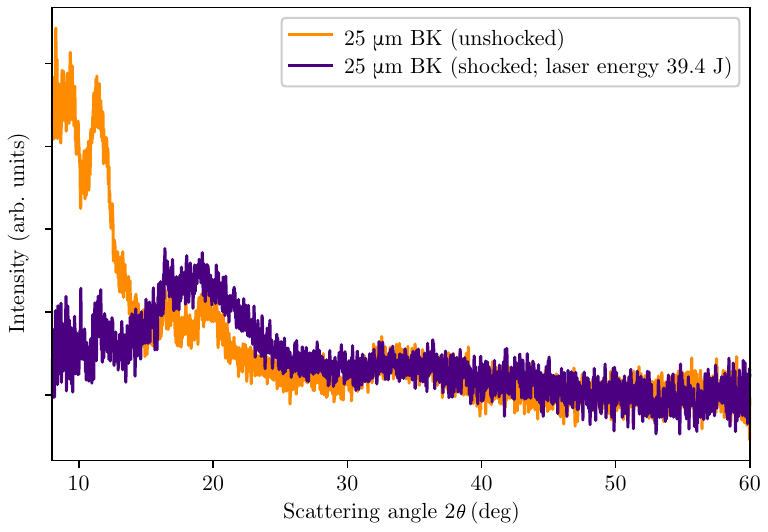} 
    \caption{A comparison of the x-ray diffraction from ambient and shock-compressed ($>100$~GPa) Kapton-B.}
    \label{F:bkcomparison}
\end{figure}

To subtract the signal due to the ablator, we used data obtained from isolated 25~$\upmu$m Kapton-B samples, both ambient and shocked with the maximum available 39.4~J for the 10~ns laser pulse and the 250~$\upmu$m phase plate, whose diffraction signals are compared in \hyperref[F:bkcomparison]{Fig.~S2}. We observed that the scattering signal is largely insensitive to shock pressure in the region $2\theta>20^\circ$ (the region in which we are primarily interested for the thermal diffuse scattering analysis) due to the Kapton's being largely structureless. We therefore used the same driven Kapton-B data for all shock pressures.

From the data shown in \hyperref[F:bkcomparison]{Fig.~S2}, we obtained the Kapton-B structure factor $S_{\textrm{BK}}$ (both ambient and shock-compressed) by dividing the raw signal through by the $I_0\times f_{\rm P} f_\Omega f_{\rm D} \times F_{\rm SA}$, where $F_{\rm SA}\approx25$~$\upmu$m. We could thus account for the different Kapton thickness used in our production samples (50~$\upmu$m).

\subsection{\label{ssec:photonflux} Beam intensity ($I_0$) correction}
\begin{figure}[t]
\includegraphics[width=1.0\columnwidth]{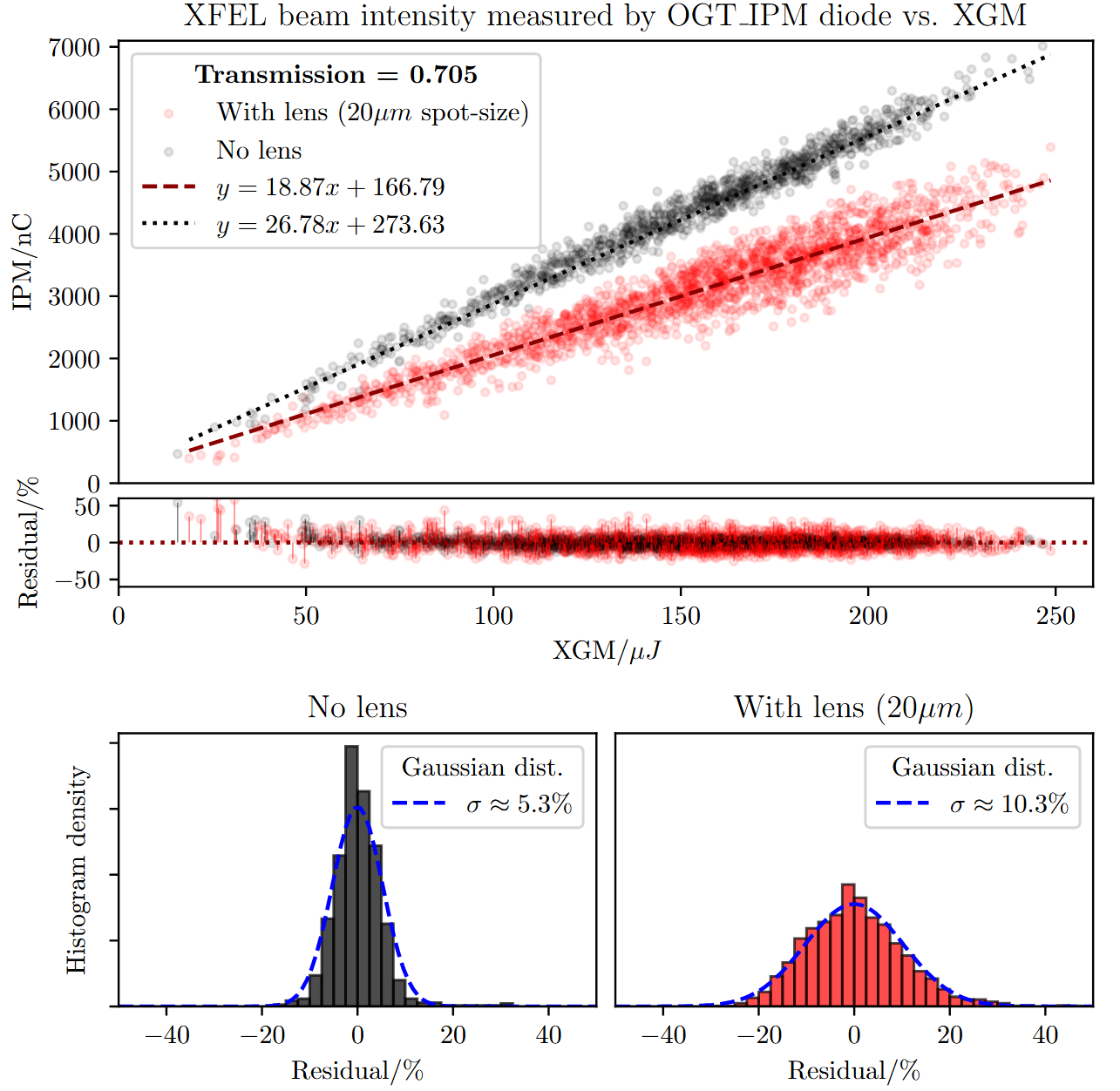} 
    \caption{Measurement of the x-ray transmission through the CRL3 lens using data collected on the XTD6 XGM (upstream of the lens) and the OGT IPM (downstream of the lens), both with and without the lens in place. 
    The above figure shows the transmission measured for the CRL3 configuration with a 20~$\upmu$m spot-size; this is repeated with the 50~$\upmu$m spot-size configuration.}
    \label{F:XGM_IPM_CRL3transmission}
\end{figure}

To account for stochastic self-amplified spontaneous emission (SASE) fluctuations in the XFEL beam intensity, $I_0$, it is essential to have a shot-to-shot x-ray intensity measure to normalize the data. Across the beamline, the European XFEL uses x-ray gas monitors (XGMs) and intensity-position monitor (IPM) diodes to measure the intensity of the beam \cite{Zastrau2021}. The XGM devices are situated upstream of the final beryllium x-ray-focusing compound reflective lenses (CRLs) (i.e., on the opposite side to the sample), and rely on the photoionization of rare gases. Their absolute measurement uncertainty is recorded as being between 7-10\%\cite{Maltezopoulos2019}, which is broadly consistent with the residuals shown in \hyperref[F:XGM_IPM_CRL3transmission]{Fig.~S3}.

The IPM diodes, on the other hand, have a much smaller measurement uncertainty and (in principle) give a more accurate on-target x-ray intensity given their closer proximity to the samples. Unfortunately, their readings were corrupted upon the firing of the DiPOLE laser, and so were unusable during this experiment. Thus, we used the XGM data to measure the x-ray intensity, correcting for the measured transmission of each spot-size used (20~$\upmu$m or 45~$\upmu$m), which required different CRL configurations.

The XGM uncertainty must be fully propagated into the extracted compressed copper structure factor, $S_{\textrm{Cu}}$, and ultimately to the temperature $T$ fitted using the Warren model. To achieve this, we simulated the effect of a noisy XGM signal by randomly selecting values $I_0$ from a normal distribution centered on the measured XGM intensity $\bar{I}_0$ with variance $\sigma=0.10$, such that every signal (Kapton only, Kapton/copper; pre-shots and main shots) was normalized by an intensity $ I_0 = \bar{I}\mathcal{N}_0(\bar{I}_0, \sigma^2)$.
 
\begin{figure}[t]
\includegraphics[width=0.95\columnwidth]{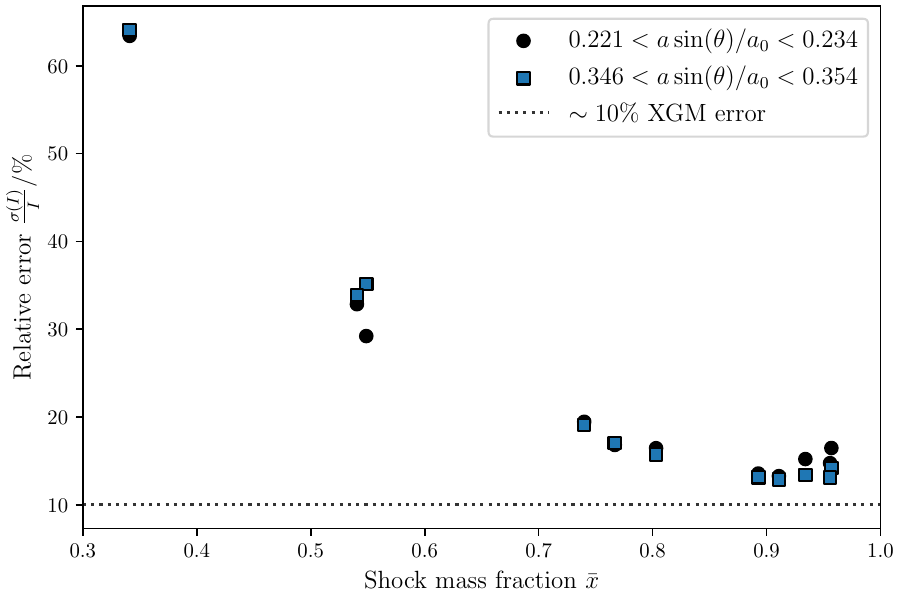} 
    \caption{The relative 68\%-percentile error $\sigma(I)$ in the inter-Bragg signal in the ranges $ 0.221 < a\sin(\theta)/a_0 <0.234 $ [between the (200) and (220) peaks] and $ 0.346 < a\sin(\theta)/a_0 < 0.354 $ [between the (222) and (400) peaks] of the extracted compressed Cu signal against the calculated mean shock fraction $\bar{x}$. The dashed horizontal line represents the ultimate $\sim10\%$ resolution limit imposed by the statistical uncertainty in the XGM reading.}
    \label{F:xerr_vs_x}
\end{figure}

Performing this procedure 1\,000 times, we obtained distributions for placing confidence limits on each quantity. \hyperref[F:xerr_vs_x]{Figure~S4} summarizes the variation in the percentage uncertainty (parametrized as the 68\% confidence interval) of the intensity $S_{\textrm{Cu}}$ in two separate inter-Bragg regions against $x$. This uncertainty is also propagated into the final fitted temperature $T$ and Debye-Waller factor $2M$ (as shown in  Figs.~8 and 9 of the Main Article).

Given the convergence for $x \gtrsim 0.8$ towards the percentage error of the XGM fluctuations (10\%), we focus on data with $x \gtrsim 0.8$ (though we note the broad agreement between the data and model even for smaller $x$).

The steps in our overall method for determining $x$ and extracting $S_{\textrm{Cu}}$ are summarized below.

\subsection{\label{ssec:method} Structure-factor isolation algorithm}
\begin{enumerate}

\item \textbf{Normalise by photon flux $I_0$}\newline
We first divide each signal (Kapton only, Kapton/copper; pre-shot and main shot) by the corresponding XGM signal. 

\item \textbf{Apply extrinsic corrective maps} \newline
We divide by non-sample related modulations $f_{\rm tot}(2\theta,\phi)$, due to x-ray polarization $f_{\rm P}$, detector solid angle $f_\Omega$, and detector nonuniformity $f_{\rm D}$.

\item \textbf{Simulate XGM statistical uncertainty}\newline
We rescale each signal by a random number $f$ with mean $\bar{f}=1$ and $\sigma=0.1$ to account for fluctuations in the XGM signal.

\item \textbf{Subtract ablator signal $I_{\rm BK}$} \newline
We subtract the Kapton-B signal, $I_{\rm BK}(2\theta,\phi)$, including the correction for x-ray attenuation by 25~$\upmu$m of downstream copper, but neglecting self-attenuation within the Kapton itself:
\begin{equation*}
    I_{\rm BK} = f_{\textrm{tot}} S_{\textrm{BK}} F_{\rm A}(\zeta|L_\textrm{Cu},\mu_{\textrm{Cu}})\ .
\end{equation*}

\item \textbf{Calculate shock fraction $x$} \newline
We find the value of $x$ which minimizes the total absolute difference between the ambient Bragg peak intensities in the pre-shot and the main shot:
\begin{equation*}
    x = \argmin_{x'\in[0,1]} \mathcal{L}(x')
\end{equation*}
where
\begin{equation*}
     \mathcal{L}(x') = \sum_{hkl} \biggr|I_{hkl}^{(\rm shot)} -  f_{\textrm{tot}} F_{\rm Cu,0}(x') S_{\textrm{Cu,0}}\biggl|
\end{equation*}
The minimization is performed using Brent's method, an iterative method using parabolic interpolation.

\item \textbf{Extract compressed copper signal} \newline
Having subtracted the ablator signal, we calculate the total copper scattering signal $I_{\rm Cu} + I_{\rm Cu,0}$ via
\begin{equation*}
\begin{split}
    I - I_{\rm BK} = f_{\rm tot}[&F_{\rm Cu}(x) S_{\rm Cu} \\ + &F_{\rm Cu,0}(x) S_{\rm Cu,0}]\ .
\end{split}
\end{equation*}
From this signal, it is possible to infer the structure factor of the compressed copper alone, $S_{\rm Cu}$, after subtraction of the appropriately weighted ambient copper structure factor, $S_{\rm Cu,0}$.

\end{enumerate}

Overall, we repeat steps 4-6 at least 1\,000 times to achieve good statistics on the effect of the XGM noise on the final intensity $S_{\rm Cu}$. The associated uncertainties are fully propagated into the temperature fitting using the model described in the next section.

\section{Warren model of elastic and inelastic x-ray scattering}

We summarize here the salient features of our thermal diffuse scattering model, the structure and development of which is owed to Warren \cite{Warren1951,Warren1953,Warren1990} and Borie \cite{Borie1961}.

\subsection{Debye-Waller factor}

Both the elastic and inelastic scattering depend on the Debye-Waller factor $M$, which parametrizes the extent to which the phonon population causes each atom to deviate from its equilibrium position. The Debye-Waller factor for an elemental crystal at temperature $T$ composed of atoms of mass $m$ is given by
\begin{equation}
    2M=  \frac{12 h^2}{m k_{\rm B}} \frac{T}{\Theta_D^2}\left [ \Phi(z) + \frac{1}{4}z  \right ] \left ( \frac{\sin \theta}{\lambda}  \right )^2 \ ,
    \label{E:debyeWaller}
\end{equation}
where $\Theta_D$ is the Debye temperature, $z = \Theta_D/T$, and where the Debye function $\Phi(z)$ is given by
\begin{equation}
    \Phi(z) = \frac{1}{z} \int_0^z dy\,\frac{y}{e^y - 1} \ .
\end{equation}
In deriving this expression for the Debye-Waller factor, several simplifications of the phonon dispersion relation must be made. First, it is assumed that each type of phonon (longitudinal or transverse) travels with the same phase velocity as the acoustic phonons of that type, regardless of wavevector. Second, there exists a maximum possible angular frequency $\omega_{m,j}$ for each type of phonon (as the crystal is discrete) whose value is parametrized by a Debye temperature $\Theta_{D,j} = \hbar\omega_{m,j}/k_B$. Under these assumptions, the Debye-Waller factor becomes
\begin{equation}
    2M = \frac{12h^2}{m k_{\rm B}}\times\frac{1}{3}\sum_{j=1}^3 \frac{T}{\Theta_{D,j}^2}\left[\Phi(z_j) + \frac{1}{4}z_j\right]\left(\frac{\sin\theta}{\lambda}\right)^2 \,
\end{equation}
where $\Theta_{D,j}$ is the Debye temperature associated with phonon mode $j$, and where $x_j = \Theta_{D,j}/T$. The final assumption is that, for every temperature, there exists an `average' Debye temperature $\Theta_D$ whose value is such that
\begin{equation}
    \frac{T}{\Theta_D^2}\left[\Phi(z) + \frac{1}{4}z\right] \approx \frac{1}{3}\sum_{j=1}^3 \frac{T}{\Theta_{D,j}^2}\left[\Phi(z_j) + \frac{1}{4}z_j\right] \ ,
\end{equation}
from which we immediately arrive at Eq.~\hyperref[E:debyeWaller]{(S21)}. An excellent exposition of the theory underpinning the Debye-Waller factor is given by James \cite{James1948}.

\subsection{Atomic form factor}
\label{ssec:atomic form factor}
To model the full scattering signal also requires that we know the atomic form factor $f$. We calculate the atomic form factor as a function of $Q \equiv q/4\pi$ via the inverse Mott-Bethe equation~\cite{Bethe1986}, expressed as a sum of $N$ Gaussian terms:
\begin{equation}
    f(Q|Z) = Z - 8 \pi^2 a_0 Q^2 \left[\alpha + \sum_{n=1}^N c_n \exp(-d_n Q^2)\right]\ ,
\end{equation}
where $a_0$ is the Bohr radius and where the atomic number $Z = 29$ for copper. We use experimental values of the form factor provided by Cromer and Mann~\cite{Cromer1968} fitted to the Mott-Bethe form using the $N=7$ updated coefficients calculated by Thorkildsen~\cite{Thorkildsen2023}.

The magnitude of the difference between $f$ and the experimental values to which it was fitted (averaged over the measured $Q$-range of $[0, 1.5]$~\AA\textsuperscript{$-1$}) is $2.2\times10^{-4} \approx 10^{-5}Z$. This error is negligible compared to other, percent-level sources of statistical error in our analysis, so Thorkildsen's $N=7$ fit is more than sufficient for the purposes of our scattering model.

\begin{figure*}
    \includegraphics{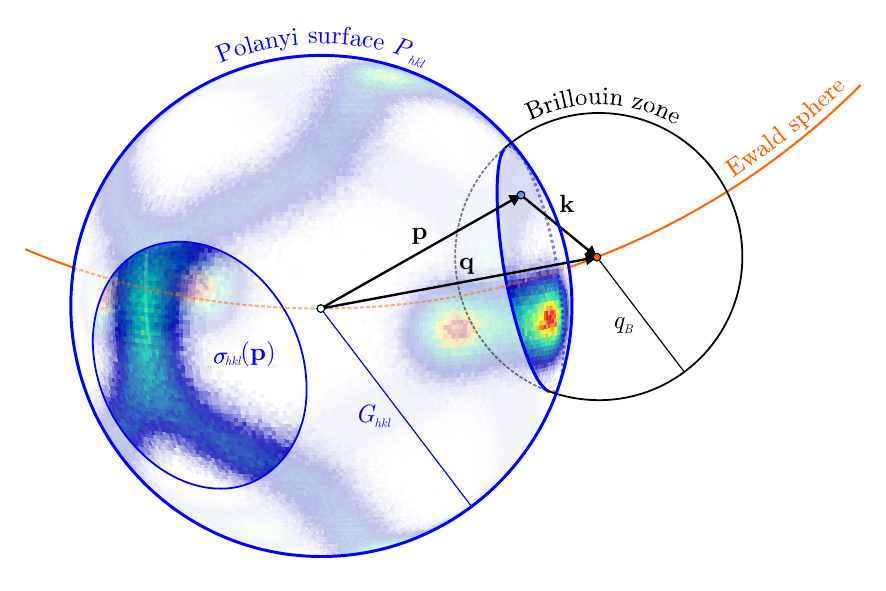} 
    \caption{Reciprocal-space construction for the calculation of thermal diffuse x-ray scattering from a polycrystal. Sphere outlined in blue depicts the Polanyi surface $P_{hkl}$ containing scattering vectors belonging to the $\{hkl\}$ family of planes, with $\sigma_{hkl}(\mathbf{p})$ scattering vectors per unit solid angle at point $\mathbf{p}$. The first-order inelastic scattering intensity at point $\mathbf{q}$ on the Ewald sphere is calculated by integrating the function $\sigma_{hkl}(\mathbf{p}) (q_B/k)^2$ over the locus of points on the Polanyi surface falling within a spherical Brillouin zone of radius $q_B$ centered on $\mathbf{q}$, with $\mathbf{k} = \mathbf{q} - \mathbf{p}$. Spherical loci are shown to scale for the $\{111\}$ Polanyi surface ($G_{111}$ = 3.01~\AA\textsuperscript{$-1$}) of an fcc crystal with lattice constant $a$ = 3.615~\AA\ ($q_B$ = 1.71~\AA\textsuperscript{$-1$}) probed by a 18~keV x-ray source ($E/\hbar c$ = 9.12~\AA\textsuperscript{$-1$}).}
    \label{F:warren_integral}
\end{figure*}

\subsection{Single-crystal scattering}

Our scattering model is built on the additive decomposition of the total x-ray scattering intensity into elastic and inelastic components owed to Warren \cite{Warren1951,Warren1953,Warren1990}. According to Warren's treatment, the contributions to the elastic and first-order inelastic scattering from a single scattering vector $\mathbf{G}$ generated by a single crystallite comprising $N_a$ atoms at reciprocal-space point $\mathbf{q}$ can be written as
\begin{equation}
\begin{split}
    s(\mathbf{q}|\mathbf{G}) ={}&s_0(\mathbf{q}|\mathbf{G}) + s_1(\mathbf{q}|\mathbf{G}) \\
    ={}&N_a f^2 e^{-2M} J(\mathbf{q} - \mathbf{G}) \\ &+ N_a f^2 2Me^{-2M} W(\mathbf{q} - \mathbf{G})\ .
    \label{E:decomposition}
\end{split}
\end{equation}
This expression omits all corrections to the measured scattering intensity per solid angle that are extrinsic to the electronic structure of the crystal, including the incident photon flux, the polarization factor, and any attenuation effects, per Eq.~\hyperref[E:master_eq]{(S9)}. The modifiers $f$ and $M$ are the atomic form and Debye-Waller factors, respectively, and the functions $J$ and $W$ describe the distribution of elastic and inelastic scattering intensity around each scattering vector $\mathbf{G}$. Here, $J$ is a sufficiently localized \textit{shape function} normalized such that $\int d^3\mathbf{k}\,J(\mathbf{k}) = 4(2\pi/a)^3$ for a face-centered cubic (fcc) crystal, and where $W$ is the Warren kernel
\begin{equation}
    W(\mathbf{k}) =
    \begin{cases} 
      \frac{1}{3}\left(\frac{q_B}{k}\right)^2 & 0 \le k \le q_B\ , \\
      0 & k > q_B\ .
   \end{cases}
\end{equation}
Here, $q_B$ is the radius of the Brillouin zone, which, in Warren's model, is assumed to be spherical and isotropic, with radius
\begin{equation}
    q_B = \frac{2\pi}{a}\left(\frac{3}{\pi}\right)^{1/3}\ .
\end{equation}
Equation~\hyperref[E:decomposition]{(S26)} above yields only the leading-order scattering from a single grain; polycrystal and higher-order scattering are discussed in Secs.~\hyperref[ssec:polycrystal]{S2D} and \hyperref[ssec:higherorder]{S2E}, respectively.

\subsection{\label{ssec:polycrystal} Polycrystal scattering}

To predict the structure factor of a full polycrystalline aggregate with a known distribution of grain orientations (i.e., a known texture) is a matter of integrating contributions from all reciprocal lattice vectors from all grains. For a polycrystal in a hydrostatic elastic strain state, the scattering vectors occupy a set of concentric spherical surfaces (\textit{Polanyi surfaces}) of radii $\{G_{hkl}\}$, where $(hkl)$ denotes the Miller indices of each family of scattering vectors. To calculate the total scattering, we perform a suitably weighted integral over each Polanyi surface (denoted by $P_{hkl}$), and subsequently sum over all surfaces:
\begin{equation}
    S(\mathbf{q}) = \sum_{hkl} \iint\limits_{P_{hkl}} d\Omega\,s(\mathbf{q}|\mathbf{p})\sigma_{hkl}(\mathbf{p})\ .
    \label{E:polanyi_integral}
\end{equation}
Here, $\sigma_{hkl}(\mathbf{p})$ is the angular density of reciprocal lattice vectors at point $\mathbf{p}$ on the Polanyi surface. For illustration, we show in \hyperref[F:warren_integral]{Fig.~S5} a representation of the integral required to calculate the contribution to the first-order thermal diffuse scattering $S_1$ from a single Polanyi surface. The Polanyi density $\sigma_{hkl}$ is in turn calculated from the polycrystal's orientation distribution function (ODF), which, for the purposes of the texture-sensitivity analysis presented in the Main Article, we took to be that of the canonical $\beta$~fiber expected of a rolled copper foil \cite{Dillamore1964,Sidor2013,Kestens2016}. We generated the corresponding ODF using \textsc{mtex}~\cite{MTEX}.

For the special case of a random powder, we can evaluate the textural integrals appearing in Eq.~\hyperref[E:polanyi_integral]{(S29)} analytically. Doing so allows us to write the following complete equations for the elastic and first-order inelastic x-ray scattering from an ideal powder sample in $q$-space:
\begin{widetext}
    \begin{align}
	S_0(\mathbf{q}) &= N_a N_g f^2 e^{-2M} \times 4\left(\frac{\sqrt{\pi}R}{a}\right)^3 \sum_{hkl} j_{hkl} \frac{1}{qG_{hkl}R^2}\exp\left[-\frac{1}{4}(q - G_{hkl})^2 R^2\right]\ ,\\
        S_1(\mathbf{q}) &= N_a N_g f^2 2Me^{-2M} \times \underbrace{\frac{1}{6} \sum_{hkl} j_{hkl} \frac{q_B^2}{qG_{hkl}} \ln\left(\frac{q_B}{|q - G_{hkl}|}\right)}_{C_1(\mathbf{q})}\ ,
        \label{E:powder_scattering}
    \end{align}
\end{widetext}
where $N_g$ is the number of grains, $j_{hkl}$ is the multiplicity of the $\{hkl\}$ peaks, and where we have assumed a Gaussian shape function (whose width is expressed via the free parameter $R$) for convenience. If we calculate the integrated intensity under a particular elastic peak with Miller indices $(hkl)$, we find that
\begin{equation}
    \int d(2\theta)\,S_{0,hkl} (2\theta) = \frac{N f^2 e^{-2M} \lambda^3 j_{hkl}}{4\pi a^3 \sin^2\theta_{hkl} \cos\theta_{hkl}}\ ,
\end{equation}
where $N = N_a N_g$ is the total number of scattering atoms, $\lambda$ is the x-ray wavelength, and $\theta_{hkl}$ is the Bragg angle at which the elastic peak is maximal. While we happen to have used a Gaussian shape function for convenience here, the result above should hold for any sufficiently localized shape function $J$. In our units, Eqs.~\hyperref[E:powder_scattering]{(S30)} and \hyperref[E:powder_scattering]{(S31)} for the elastic and first-order inelastic scattering are identical to those derived by Warren.

\subsection{\label{ssec:higherorder} Higher-order scattering}

Equation \hyperref[E:decomposition]{(S26)} gives only the leading-order contribution to the inelastic scattering from a finite-temperature polycrystal: there exist higher-order, diminishing contributions to the scattering from multi-phonon processes. In general, the thermal diffuse scattering may be expressed as a sum of $\ell$-phonon scattering events, which takes the form of a power series in $2M$:
\begin{align}
    S_{\textrm{TD}}(\mathbf{q}) &= \sum_{\ell=1}^\infty S_{\ell}(\mathbf{q}) \\
    &= N f^2 e^{-2M} \sum_{\ell=1}^\infty \frac{(2M)^\ell}{\ell!}C_{\ell}(\mathbf{q})\ ,
    \label{E:higher_order}
\end{align}
where $C_{\ell}$ encodes the structure of the $\ell$\textsuperscript{th}-order diffuse scattering. Calculation of the coefficients $\{C_{\ell}\}$ is possible within the Debye-crystal framework used by Warren, but quickly complexifies with increasing $\ell$. We use an accurate approximation for the all-order thermal diffuse scattering owed to Borie \cite{Borie1961} that exploits the decaying structure of $C_{\ell}$ at higher orders. He observed that, to a reasonable approximation,
\begin{align}
    C_{\ell} =
    \begin{cases} 
      \frac{1}{2}(1 + C_1) & \ell = 2\ , \\
      1 & \ell \ge 2\ ,
   \end{cases}
\end{align}
where $C_1$ [defined in Eq.~\hyperref[E:powder_scattering]{(S31)}] describes the structure of the first-order inelastic scattering $S_1$. Substituting the coefficients above into Eq.~\hyperref[E:higher_order]{(S34)} yields Borie's approximation for the all-order thermal diffuse scattering:
\begin{equation}
\begin{split}
    S_{\mathrm{TD}}(\mathbf{q}) ={}&N f^2 (1 - e^{-2M}) \\ &+ N f^2 e^{-2M}(2M + M^2)(C_1 - 1)\ .
\end{split}
\end{equation}
This is the expression we use to treat the experimental data in the Main Article. Without the contributions from the higher-order scattering, we find that we systematically underestimate the diffuse scattering observed at higher values of $2\theta$.

\section*{References}
%